\documentclass{aa}

\usepackage{wasysym}

\usepackage{graphicx}
\usepackage{amsmath}   
\usepackage{tabulary}
\usepackage{epsfig}
\usepackage{amssymb}
\usepackage{multirow}
\usepackage{appendix}
\usepackage{natbib}
\usepackage{lineno}
\usepackage{wasysym}
\usepackage{color}         
\usepackage{enumerate}
\usepackage{tikz}
\usepackage{esvect}       

\usepackage{graphicx,url,twoopt,natbib}
\usepackage{txfonts}
\usepackage[pdftitle={Low-cost precursor of an interstellar mission}, colorlinks = true, breaklinks = true, citecolor = blue, linkcolor = blue, urlcolor = blue, pdfauthor = {Ren\'{e} Heller}]{hyperref}
\bibpunct{(}{)}{;}{a}{}{,}    

\begin{document} 

\title{Low-cost precursor of an interstellar mission}

\titlerunning{Low-cost precursor of an interstellar mission}

\author{Ren\'{e} Heller\inst{1}
             \and
             Guillem Anglada-Escud\'e\inst{2,3}
             \and
             Michael Hippke\inst{4,5}
             \and
             Pierre Kervella\inst{6}
}

\institute{
Max Planck Institute for Solar System Research, Justus-von-Liebig-Weg 3, 37077 G\"ottingen,
Germany, \href{mailto:heller@mps.mpg.de}{heller@mps.mpg.de}
\and
Institut de Ci{\`e}ncies de l’Espai, Consejo Superior de Investigaciones Cient{\'i}ficas, Campus Universitat Aut{\`o}noma de Barcelona, 08193 Bellaterra, Spain \href{mailto:guillem.anglada@gmail.com}{guillem.anglada@gmail.com}
\and
Institut d'Estudis Espacials de Catalunya, Edifici Nexus, Desp. 201, 08034 Barcelona, Spain
\and
Sonneberg Observatory, Sternwartestr. 32, 96515 Sonneberg, Germany, \href{mailto:michael@hippke.org}{michael@hippke.org}
\and
Visiting Scholar, Breakthrough Listen Group, Berkeley SETI Research Center, Astronomy Department, UC Berkeley
\and
LESIA, Observatoire de Paris, Universit\'e PSL, CNRS, Sorbonne Universit\'e, Universit\'e de Paris, 5 place Jules Janssen, 92195 Meudon, France
}
\date{Received 18 June 2020; Accepted 7 July 2020}
 
\abstract{
The solar photon pressure provides a viable source of thrust for spacecraft in the solar system. Theoretically it could also enable interstellar missions, but an extremely small mass per cross section area is required to overcome the solar gravity. We identify aerographite, a synthetic carbon-based foam with a density of $0.18\,{\rm kg\,m}^{-3}$ (15,000 times more lightweight than aluminum) as a versatile material for highly efficient propulsion with sunlight. A hollow aerographite sphere with a shell thickness $\epsilon_{\rm shl}~=~1$\,mm could go interstellar upon submission to  solar radiation in interplanetary space.
Upon launch at 1\,AU from the Sun, an aerographite shell with $\epsilon_{\rm shl}~=~0.5$\,mm arrives at the orbit of Mars in 60\,d and at Pluto's orbit in 4.3\,yr. Release of an aerographite hollow sphere, whose shell is $1\,\mu$m thick, at $0.04$\,AU (the closest approach of the Parker Solar Probe) results in an escape speed of nearly $6900\,{\rm km\,s}^{-1}$ and 185\,yr of travel to the distance of our nearest star, Proxima\,Centauri. The infrared signature of a meter-sized aerographite sail could be observed with JWST up to 2\,AU from the Sun, beyond the orbit of Mars. An  aerographite hollow sphere, whose shell  is $100\,\mu$m thick, of  1\,m (5\,m) radius weighs 230\,mg (5.7\,g) and has a $2.2$\,g (55\,g) mass margin to allow interstellar escape. The payload margin is ten times the mass of the spacecraft, whereas the payload on chemical interstellar rockets is typically   a thousandth of the weight of the rocket. Using 1\,g (10\,g) of this margin (e.g., for miniature communication technology with Earth), it would reach the orbit of Pluto 4.7\,yr (2.8\,yr) after interplanetary launch at 1\,AU.
Simplistic communication would enable studies of the interplanetary medium and a search for the suspected Planet Nine, and would serve as a precursor mission to $\alpha$\,Centauri. We estimate prototype developments costs of 1\,million USD, a price of 1000\,USD per sail, and a total of $<~10$\,million USD including launch for a piggyback concept with an interplanetary mission.
}


\maketitle

\section{Introduction}

The discovery of a roughly Earth-mass planet candidate in the habitable zone \citep{1993Icar..101..108K} around our nearest stellar neighbor \object{Proxima Centauri} (Proxima\,Cen) \citep{2016Natur.536..437A} and recent evidence of a Neptune-mass planet candidate \citep{2020SciA....6.7467D,2020A&A...635L..14K,2020RNAAS...4...46B} motivates a reconsideration of the possibility of interstellar travel. Chemically driven rockets are not suited for interstellar exploration on a timescale comparable to the human life span, due to their fundamental limitations rooted in the rocket equation \citep{tsiolkovsky1903}. The Voyager~1 spacecraft, at a speed of about $17\,{\rm km\,s}^{-1}$ or roughly $3.6\,{\rm AU\,yr}^{-1}$ (the fastest of five vehicles that ever acquired escape speed from the solar system), would reach the nearest star, Proxima\,Cen at a distance of $4.2439\,\pm\,0.0012$\,light-years \citep[at epoch 2015.5;][]{2018A&A...616A...1G} in about 75,000\,yr.

Instead, interstellar speeds could be achieved by shooting an extremely powerful ground-based laser beam at a light sail in space \citep{1966Natur.211...22M}. For a nominal 10\,m$^2$ sail with a mass per cross section ratio of $\sigma~=~0.1\,{\rm kg\,m}^{-2}$ the required power would be ${\sim}1$\,TW \citep{1967Natur.213..588R}. The Breakthrough Starshot Initiative\footnote{\href{http://breakthroughinitiatives.org}{http://breakthroughinitiatives.org}} has been investigating this possibility of using laser technology to accelerate highly reflective light sails to interstellar speeds. Some of the key challenges of this concept are in the extreme power output required during the launch phase \citep[10-100\,GW for several minutes;][]{2016JBIS...69...40L,Kulkarni_2018}, the stability of the sail riding on a collimated laser beam in the presence of atmospheric perturbations \citep{2017ApJ...837L..20M}, the extreme accelerations of $\sim10^4\,g$ ($g$ being the Earth's surface gravity) acting upon a proposed one-gram sail during launch \citep{2016JBIS...69...40L}, the aiming precision towards the target star \citep{2017AJ....154..115H}, and the structural integrity of the sail while being heated to 1000\,K or more during launch \citep{Atwater2018}.

Alternatively, the solar irradiation could be used to propel ultra thin and ultra lightweight sails to interstellar speeds \citep{Cassenti1997} and even allow deceleration at their target star systems \citep{Matloff2009,2017ApJ...835L..32H,2017AJ....154..115H}. Closer to home, several solar sail missions have already demonstrated the feasibility of using sunlight as a thrust. The Light Sail 2 mission\footnote{\href{http://www.planetary.org/explore/projects/lightsail-solar-sailing}{www.planetary.org/explore/projects/lightsail-solar-sailing/}} successfully performed controlled solar sailing \citep{doi:10.2514/6.2020-2177} in low Earth orbit (LEO). Prior to the Light Sail project, the IKAROS mission successfully performed acceleration and attitude control using its solar sail during a six-month voyage to Venus in 2010 \citep{2013AcAau..82..183T}. Dedicated reflectivity changes in its 80 liquid crystal panels were used to torque the sail using solar photons alone. 
IKAROS \citep{2011AcAau..69..833T} had a square format with a 20\,m diagonal, and was made of a 7.5\,$\mu$m thick sheet of polyimide. The polyimide sheet had an areal mass density of about $10\,{\rm g\,m}^{-2}$, resulting in a total sail mass of 2\,kg. While this setup allowed IKAROS to gain about $400\,{\rm m\,s}^{-1}$ of speed from the Sun within almost three years of operation, it is impossible for the solar photon pressure to accelerate such a sail to interstellar speeds \citep{2017ApJ...835L..32H}.

Graphene has been suggested as a candidate material for a Sun-driven, interstellar photon sail, due to its extremely low mass per cross section ratio \citep[$\sigma~=~7.6\times10^{-7}\,{\rm kg\,m}^{-2}$;][]{Peigney2001507}. In theory, a graphene-based sail could achieve high velocities \citep{2012JBIS...65..378M,2013JBIS...66..377M}. As a principal caveat, however, graphene is almost completely transparent to optical light with a reflectivity close to zero ($\mathcal{R}=0$) and an absorptivity of just about 2.3\,\% ($\mathcal{A}=0.023$)
\citep{Nair1308}. As a consequence, its transmissivity $\mathcal{T}=0.977$, because $\mathcal{A} + \mathcal{R} + \mathcal{T}=1$. The absorptive and reflective properties of graphene can be greatly enhanced by doping graphene monolayers with alkali metals \citep{Jung2011} or by sandwiching them between substrates \citep{Yan2012}, but this comes at the price of greatly increasing $\sigma$. The limited structural integrity of a graphene monolayer requires additional material thereby further increasing $\sigma$ and complicating the experimental realization. All of this ultimately ruins the beautiful theory of a pure graphene sail.

In this work we present a new concept that avoids many of the above-mentioned obstacles that could serve as a low-cost precursor to an interstellar mission. Our concept involves a hollow sphere, approximately one meter in diameter, made of aerographite (our ``solar sail''), which is first brought to space (LEO, translunar orbit, or interplanetary space) by a conventional rocket and then released to the solar photon pressure for acceleration to interstellar speed.

\section{Interstellar escape from interplanetary space}

\subsection{Critical mass per cross section}

We start by deriving the condition for a solar sail to become unbound from the gravitational attraction of the Sun, which is met if the total force ($F_{\rm tot}$) on the sail is positive for arbitrary distances to the Sun. For now we assume that the sail is at a radial distance $r$ from the Sun and sufficiently far from any planetary gravitational field, that is, in interplanetary space. Neglecting possible effects from the solar wind as well as any drag force from the interplanetary medium, we consider that the total force is composed of the repulsive force due to the solar radiation \citep[$F_{\rm rad}$;][]{1979Icar...40....1B} and the attractive gravitational force between the sail and the Sun ($F_{\rm grav}$),

\begin{align}
F_{\rm rad} & = \frac{1}{c} \frac{L_\odot}{4\pi r^2} \, S \kappa_{\rm rad} \label{eq:F_rad} \ ,\\
F_{\rm grav} & = -\, \frac{G M_\odot}{r^2} \,m \hspace{0.35cm} \ ,
\end{align}

\noindent
where $c$ is the speed of light, $G$ the gravitational constant, $L_\odot$ the solar luminosity, $M_\odot$ the solar mass, $S$ the cross sectional area of the sail presented to the solar radiation, $m$ the mass of the sail, and $\kappa_{\rm rad}~=~\mathcal{A} + 2\mathcal{R}$  the radiation pressure coupling efficiency of the sail, which depends on the absorptive--reflective properties of the sail material. Usually, $\kappa_{\rm rad}~=~1$ but for a fully transparent material $\kappa_{\rm rad}~=~0$, whereas for a fully reflective material and perpendicular reflection $\kappa_{\rm rad}~=~2$. Moreover, all fully opaque objects have $\kappa_{\rm rad}~\geq~1$.

Equation~\eqref{eq:F_rad} assumes that all wavelengths of the solar spectral energy distribution are absorbed or reflected to the same extent. This neglect of the wavelength-dependence of the sail material is lifted in Sect.~\ref{sec:kappa}. For now, the total force appears as

\begin{eqnarray} \label{eq:totalforce}
F_{\rm tot} = F_{\rm rad} +F_{\rm grav} = \frac{1}{r^2}\left(\frac{L_\odot}{4\pi c} \, S \kappa_{\rm rad} - G M_\odot m\right) \ .
\end{eqnarray}

\noindent
The mass per cross section area of the sail is $\sigma~=~m/S$, which can be substituted in the right-hand side of Eq.~\eqref{eq:totalforce} as

\begin{eqnarray} \label{eq:totalforce2}
F_{\rm tot} = \frac{1}{r^2}\left(\frac{L_\odot}{4\pi c} \, S \kappa_{\rm rad} - G M_\odot \ \sigma S \right) \ .
\end{eqnarray}

\noindent
The condition for the sail to become unbound from the solar system is $F_{\rm tot}~>~0$, which leads to a condition that is independent of $r$,

\begin{eqnarray}
\frac{L_\odot}{4\pi c} \, \kappa_{\rm rad} - G M_\odot \sigma > 0 \label{eq:condition} \,
,\end{eqnarray}

\noindent
or equivalently

\begin{eqnarray} \label{eq:conditionSigma}
\frac{\sigma}{\kappa_{\rm rad}} < \frac{L_\odot}{4\pi c G M_\odot}  \ .
\end{eqnarray}

\noindent
Obviously, the right-hand side of Eq.~\eqref{eq:conditionSigma} has units of an areal mass density (kg\,m$^{-2}$), which we define as $\sigma_\odot$. This critical value is the surface density required for a solar system\footnote{Other star systems have their own value of  critical surface density for objects to become interstellar (see Appendix~\ref{sec:app_other}).} object to become interstellar,

\begin{eqnarray} \label{eq:sigma_I}
\sigma_\odot = \frac{L_\odot}{4\pi c G M_\odot} = 7.6946 \times 10^{-4}\,{\rm kg\,m}^{-2} \ .
\end{eqnarray}

\noindent
This value is about three orders of magnitude higher (i.e.,  more tolerant than the material constant of graphene), suggesting that it should be possible to construct Sun-driven interstellar sails with more conventional materials and possibly with weight margins for onboard instrumentation. This insight, given by the value of Eq.~\eqref{eq:sigma_I}, is the key to the mission concept proposed in this paper.

\subsection{Sail designs}

\subsubsection{Filled cuboid}
\label{sec:cuboid}

As a first approach towards estimating plausible physical dimensions of a sail and to identify feasible sail materials and designs, we start by considering a cuboid shape with mass per volume density $\rho$. Its mass is given as $m~=~{\rho}lS$. We assume that its front side offers an effective cross section $S$ towards the Sun and that it has a length (or thickness) $l$ between its front and back sides. Then

\begin{eqnarray}
\sigma_{\rm cub} = \frac{m}{S} = \frac{\rho l S}{S} = \rho l \ ,
\end{eqnarray}

\noindent
which only depends on the thickness of any given material. By substituting $\rho l$ for $\sigma_\odot$ in the left-hand side  of Eq.~\eqref{eq:conditionSigma}, we derive the critical value for the thickness of a cuboid box of material to become interstellar from interplanetary space as

\begin{eqnarray} \label{eq:l_int}
l_{\rm cub} < \frac{\kappa_{\rm rad}}{\rho} \sigma_\odot \equiv l_{\rm cub}'
.\end{eqnarray}

\subsubsection{Filled sphere}

Moving on to a solid sphere of radius $l$ and cross section area $\pi l^2$ as an alternative geometry of a sail, we have

\begin{eqnarray}
\sigma_{\rm sph} = \frac{m}{S} = \frac{\rho 4/3 \pi l^3}{\pi l^2} = \frac{4}{3} \rho  l  \ .
\end{eqnarray}

\noindent
Substitution of $4{\rho}l/3$ for $\sigma$ in the  left-hand side of Eq.~\eqref{eq:conditionSigma} yields

\begin{eqnarray} \label{eq:l_sph}
l_{\rm sph} < \frac{3}{4} \frac{\kappa_{\rm rad}}{\rho} \sigma_\odot \equiv l_{\rm sph}'
\end{eqnarray}

\noindent
for the critical thickness of a spherical sail in interplanetary space to become interstellar.

 \begin{figure*}[t]
\centering
\includegraphics[angle= 0, width=.495\linewidth]{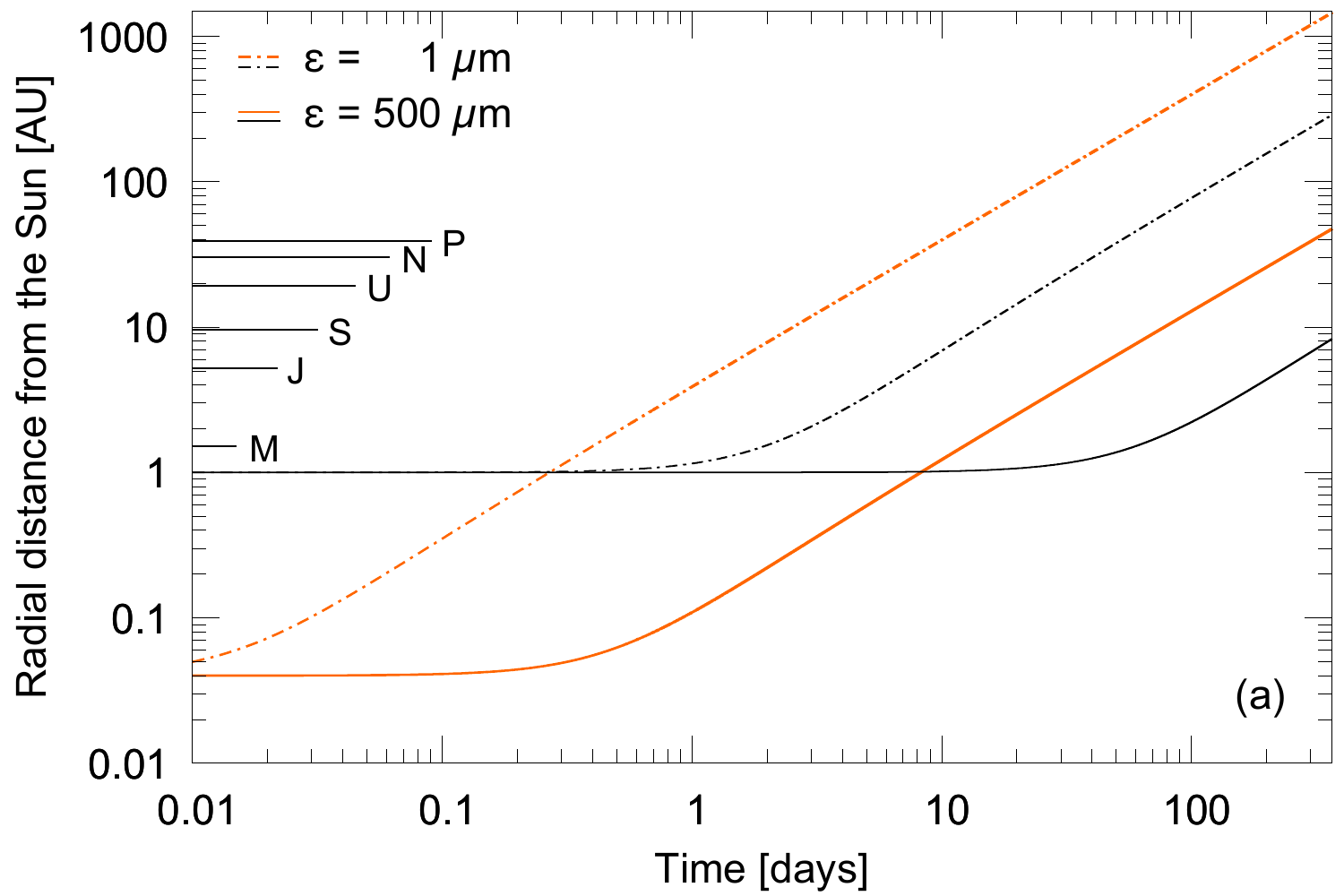}
\includegraphics[angle= 0, width=.495\linewidth]{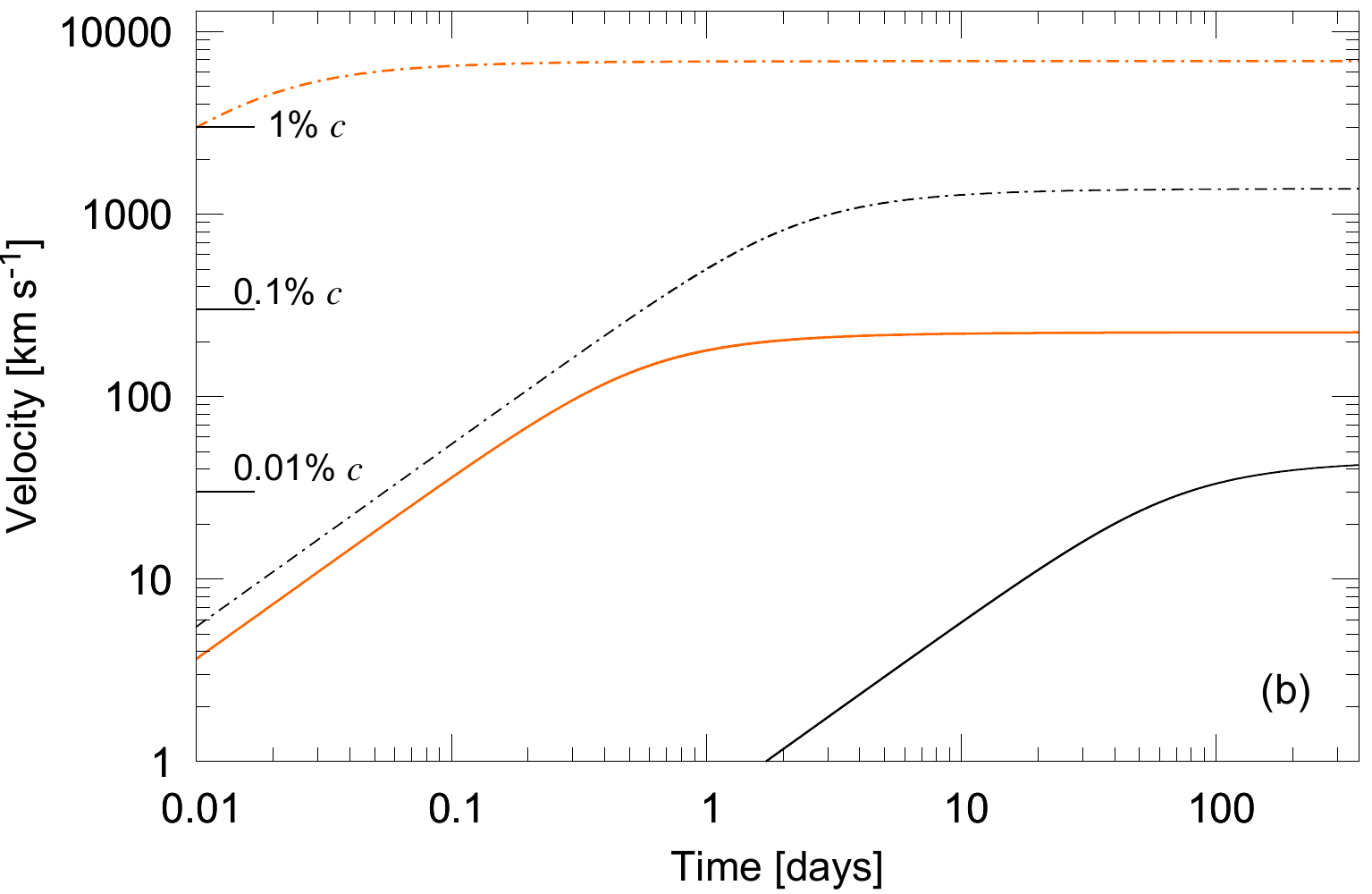}
\caption{Travel characteristics of a solar radiation driven aerographite ($\rho~=~0.18\,{\rm kg\,m}^{-3}$) hollow sphere with a shell thickness $\epsilon$. Tracks were computed through numerical integration of the total in force Eq.~\eqref{eq:totalforce} divided by the sail mass, which equals the sail acceleration. Orange lines refer to launch at 0.04\,AU from the Sun. Black lines refer to launch from interplanetary space (Earth's gravity being negligible) at 1\,AU. (a) Radial distance of the sail from the Sun as a function of time. The orbits of the solar system outer planets and of Pluto are indicated by their initials. (b) Radial velocity of the sail with respect to the Sun as a function of time.}
\label{fig:travel_ss}
\end{figure*}

\subsubsection{Hollow sphere (or shell)}
\label{sec:sphere}

The mass per cross section ratio of a spherical sail design can be decreased substantially if the sphere is hollow. If we consider a shell with an outer radius $l$ and a thickness $\epsilon$, we find

\begin{align}\nonumber
\sigma_{\rm shl} &= \frac{m}{S} = \frac{\rho}{\pi l^2} {\Bigg (} \frac{4 \pi}{3} l^3 - \frac{4 \pi}{3}(l-\epsilon)^3 {\Bigg )}\\ \nonumber
                 &= \frac{4}{3}\rho \left( 3 l^2 \epsilon - 3 l \epsilon^2 + \epsilon^3 \right) = 4 \rho \epsilon + \mathcal(O)\left(\frac{\epsilon^2}{l}\right)\\
                 &\approx 4 \rho \epsilon \ ,
                 \label{eq:shell}
\end{align}

\noindent
where the last line is accurate to $<1\,\%$ if the thickness of the shell is less than a tenth of the shell radius, $\epsilon < l/10$. Substitution of $4{\rho}\epsilon$ for $\sigma$ in the left-hand side of Eq.~\eqref{eq:conditionSigma} yields

\begin{eqnarray} \label{eq:eps_shl}
\epsilon_{\rm shl} < \frac{1}{4} \frac{\kappa_{\rm rad}}{\rho} \sigma_\odot \equiv \epsilon_{\rm shl}'
\end{eqnarray}

\noindent
for the critical thickness of a shell-like sail in interplanetary space to become interstellar.

\subsubsection{Material and design of solar interstellar sails}
\label{sec:design}

A comparison of Eqs.~\eqref{eq:l_int}, \eqref{eq:l_sph}, and \eqref{eq:eps_shl} reveals that generally the maximum dimension for a structure to become interstellar by the solar photon pressure is  on the order of $l_{\rm max}'~{\approx}~\kappa_{\rm rad}\sigma_\odot / \rho$. If we approximate $\kappa_{\rm rad}=1$ for now and assume that the material has a density similar to that of solid carbon ($\rho=2260\,{\rm kg\,m}^{-3}$), then we can derive an estimate of $l_{\rm max}'~=~340\,{\rm nm}$. For comparison, if we consider an ultra lightweight material such as aerographite, which has been demonstrated to exhibit an extremely low density near $0.18\,{\rm kg\,m}^{-3}$ \citep{Mecklenburg2012}, we obtain $l_{\rm max}'~=~4.27$\,mm.

In Table~\ref{tab:materials} we summarize our estimates of $l_{\rm max}'$ for a range of selected materials. In particular, for aluminum we assumed canonical values of $\rho~=~2700\,{\rm kg\,m}^{-3}$ and $\kappa_{\rm rad}~=~1.8$, and for Mylar film we used $\rho~=~1390\,{\rm kg\,m}^{-3}$ and $\kappa_{\rm rad}~=~1.9$. Both aluminum foil and Mylar film are very reflective. In addition to aerographite we also studied carbon nanofoam as an alternative ultra lightweight material with $\rho~=~2\,{\rm kg\,m}^{-3}$ and $\kappa_{\rm rad}~\sim~1$ \citep{Rode2000}. Among all the materials listed in Table~\ref{tab:materials}, aerographite particularly stands out.

With respect to the shape of a solar sail, aerographite as a sail material implies a maximum edge length of $4.27$\,mm for a cube to become interstellar (Eq.~\ref{eq:l_int}). For a filled aerographite sphere, Eq.~\eqref{eq:l_sph} means a radius of $3.21$\,mm for the object to become interstellar. These values demonstrate that the dimensions of such a vehicle would   be limited to scales that are  too small to be useful. But for a hollow sphere the critical length is the thickness $\epsilon$, and not the radius of the shell. To leading order, Eq.~\eqref{eq:eps_shl} is independent of the shell radius, which means that a hollow sphere can virtually have an arbitrarily large radius as long as the thickness of the shell is $\epsilon_{\rm shl}~\lesssim~1/4 \times 4.27\,{\rm mm}~=~1.07\,{\rm mm}$ for an aerographite-based sail. In the following we use $\epsilon_{\rm shl,aer}'~=~1$\,mm as the critical thickness of a hollow aerographite sphere in interplanetary space to become gravitationally unbound from the solar system due to the solar photon pressure.

\subsection{Geometrical and absorptive--reflective coupling}
\label{sec:kappa}

Equations~\eqref{eq:l_int} and \eqref{eq:l_sph} reveal a fundamental relationship between the critical length scale of a sail with arbitrary geometry ($l_{\rm arb}$), its geometric--radiative coupling, and the density of its material as

\begin{equation}
l_{\rm arb} < \kappa_{\rm geo} \frac{\kappa_{\rm rad}}{\rho} \sigma_\odot \ ,
\end{equation}

\noindent
where $\kappa_{\rm geo}$ is a geometric coupling constant for the incoming radiation. In particular, for a cuboid we find $\kappa_{\rm geo,cub}=1$, for a sphere $\kappa_{\rm geo,sph}=3/4$, and for a shell $\kappa_{\rm geo,sph}=1/4$.

In general, the reflective--absorptive properties of any material are functions of the wavelength ($\lambda$). As a consequence, $\kappa_{\rm rad}$ is obtained by integrating the reflection and absorption coefficients of the material over the relevant bandwidth of the incoming radiation. Most of the photonic energy of the Sun is emitted within $200\,{\rm nm}~\lesssim~\lambda~\lesssim~1\,\mu$m.

\begin{table}
\caption{Example materials for light sails.}
\label{tab:materials}
\centering
\begin{tabular}{cccc}
\hline\hline
Material & $\rho$ [kg\,m$^3$] & $\kappa_{\rm rad}$ & $l_{\rm max}'$ \\
\hline
Aerographite & 0.18 & ${\sim}~1$ & $4.27~{\times}~10^{-3}$\,m \\
Carbon nanofoam & 2 & ${\sim}~1$ & $3.85~{\times}~10^{-4}$\,m \\ 
Mylar film & 1390 & ${\sim}~1.9$ & $1.05~{\times}~10^{-6}$\,m \\ 
Aluminum foil & 2700 & ${\sim}~1.8$ & $5.13~{\times}~10^{-7}$\,m \\ 
Sand (SiO$_2$) & 2600 & ${\sim}~1$ & $2.96~{\times}~10^{-7}$\,m \\ 
\hline
\end{tabular}\\
\tablefoot{Columns give typical mass volume densities ($\rho$), radiation coupling constants ($\kappa_{\rm rad}$), and resulting characteristic length (or thickness) as $\sigma_\odot \kappa_{\rm rad}/\rho$. Detailed properties of aerographite were described by \citet{Mecklenburg2012}, those of carbon nanofoam by \citet{Rode2000}.}
\end{table}

\subsection{Numerical integration of the force equation}
\label{sec:numerical_1D}

Equipped with the necessary expressions for a given sail shape and composition, we can now compute 1D trajectories of a sail through the solar system. We divide $F_{\rm tot}(r)$ from Eq.~\eqref{eq:totalforce} by the sail mass, which provides us with the sail radial acceleration, $a~=~F_{\rm tot}(r)_{t}/m$ at time $t$. Then we integrate the equations of motion numerically for one year of simulated time using a constant time step (${\Delta}t$) of one minute:

\begin{align}
r(t + {\Delta}t) =& \ r(t) + {\Delta}t \frac{{\rm d}r}{{\rm d}t} = r(t) + {\Delta}t \ v(t) \\
v(t + {\Delta}t) =& \ v(t) + {\Delta}t \frac{{\rm d}v}{{\rm d}t} = v(t) + {\Delta}t \frac{F_{\rm tot}(r)_{t}}{m} \ .
\end{align}

\noindent
For all trajectories we assumed zero initial velocity, $v(t=0)=0$.

In Fig.~\ref{fig:travel_ss} we present the tracks resulting from these numerical integrations. Black lines refer to a sail launch at 0.04\,AU from the Sun, which is the closest approach of the Parker Solar Probe \citep[formerly the Solar Probe Plus Mission;][]{2016SSRv..204....7F}. Red lines refer to a launch at 1\,AU from the Sun, but well outside the Earth's gravitational potential (Eq.~\ref{eq:totalforce} ignores the effect of the Earth). For the sail material and shape we assume an aerographite hollow sphere, for which there is $\sigma~=~\sigma_{\rm shl}~=~4\rho\epsilon$ (Eq.~\ref{eq:shell}) and $\rho~=~0.18\,{\rm kg\,m}^{-3}$. Solid lines use a shell thickness of $\epsilon~=~0.5\,\epsilon_{\rm shl,aer}'~=~500\,\mu$m and dash-dotted lines use $\epsilon~=~0.001\,\epsilon_{\rm shl,Cnf}'~=~1\,\mu{\rm m}$. The radiative coupling is set to $\kappa_{\rm rad}~=~1$.

In Fig.~\ref{fig:travel_ss}(a) we plot $r(t)$. A hollow aerographite sail with $\epsilon~=~1\,\mu$m deployed at 0.04\,AU from the Sun reaches the orbit of Mars in about 0.4\,d or roughly 10\,hr and the orbit of Pluto within 9.9\,d. The same sail with $\epsilon~=~1\,\mu$m but launched from 1\,AU takes 2\,d to the orbit of Mars and 52\,d to the orbit of Pluto. A relatively thick aerographite hollow sphere with $\epsilon~=~500\,\mu$m takes 12\,d to the orbit of Mars and 304\,d to the orbit of Pluto if launched at 0.04\,AU from the Sun. For comparison, if launched from 1\,AU it would take 60\,d and 4.2\,yr to the orbits of Mars and Pluto, respectively.

Figure~\ref{fig:travel_ss}(b) illustrates $v(t)$ for the same sail properties as used in panel (a). In particular, this plot shows that thinner spheres (see dash-dotted lines), or more generally sails with a lower $\sigma/\kappa_{\rm rad}$ value, achieve their terminal speeds relatively quickly. The effect increases the closer the sail starts to the Sun. The reason for this is  their fast escape to large distances where the solar flux is extremely weak. The hollow sphere with  $1\,\mu$m shell simulated to launch at 0.04\,AU reaches a speed of 1\%\,$c$ within 0.01\,d or 14.4\,min and a terminal speed of 2.3\%\,$c$ within about 20\,d. The maximum acceleration occurs at launch and is about $400\,g$. To the contrary, thicker spheres (solid lines) continue to be accelerated for much longer in particular if they start at greater solar distances. The $\epsilon~=~500\,\mu$m hollow sphere approaches its terminal speed near 42\,km\,s$^{-1}$ after about a year. Its maximum acceleration is $6.9~\times~10^{-4}\,g$.

 \begin{figure*}[t]
\centering
\includegraphics[angle= 0, width=.495\linewidth]{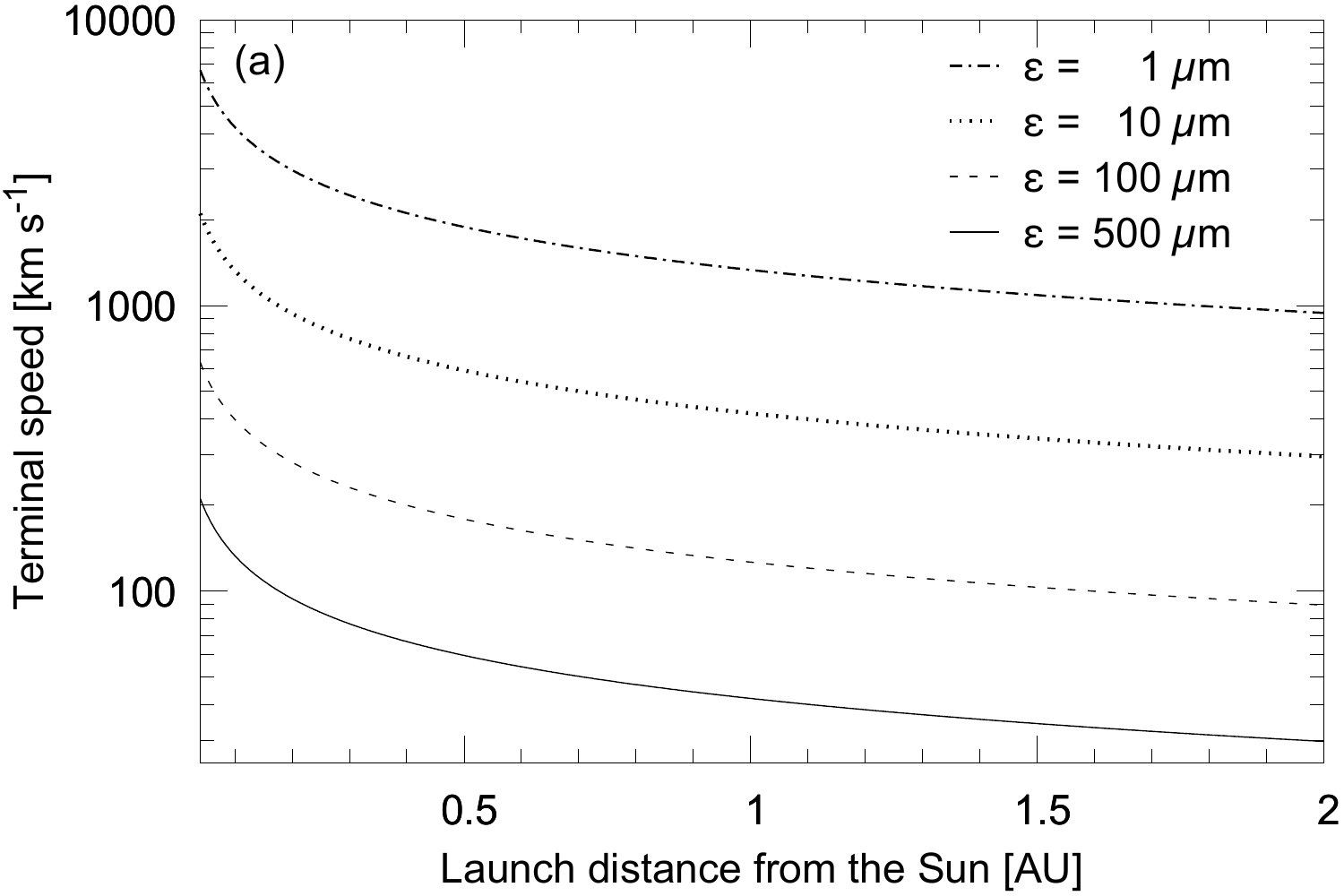}
\includegraphics[angle= 0, width=.495\linewidth]{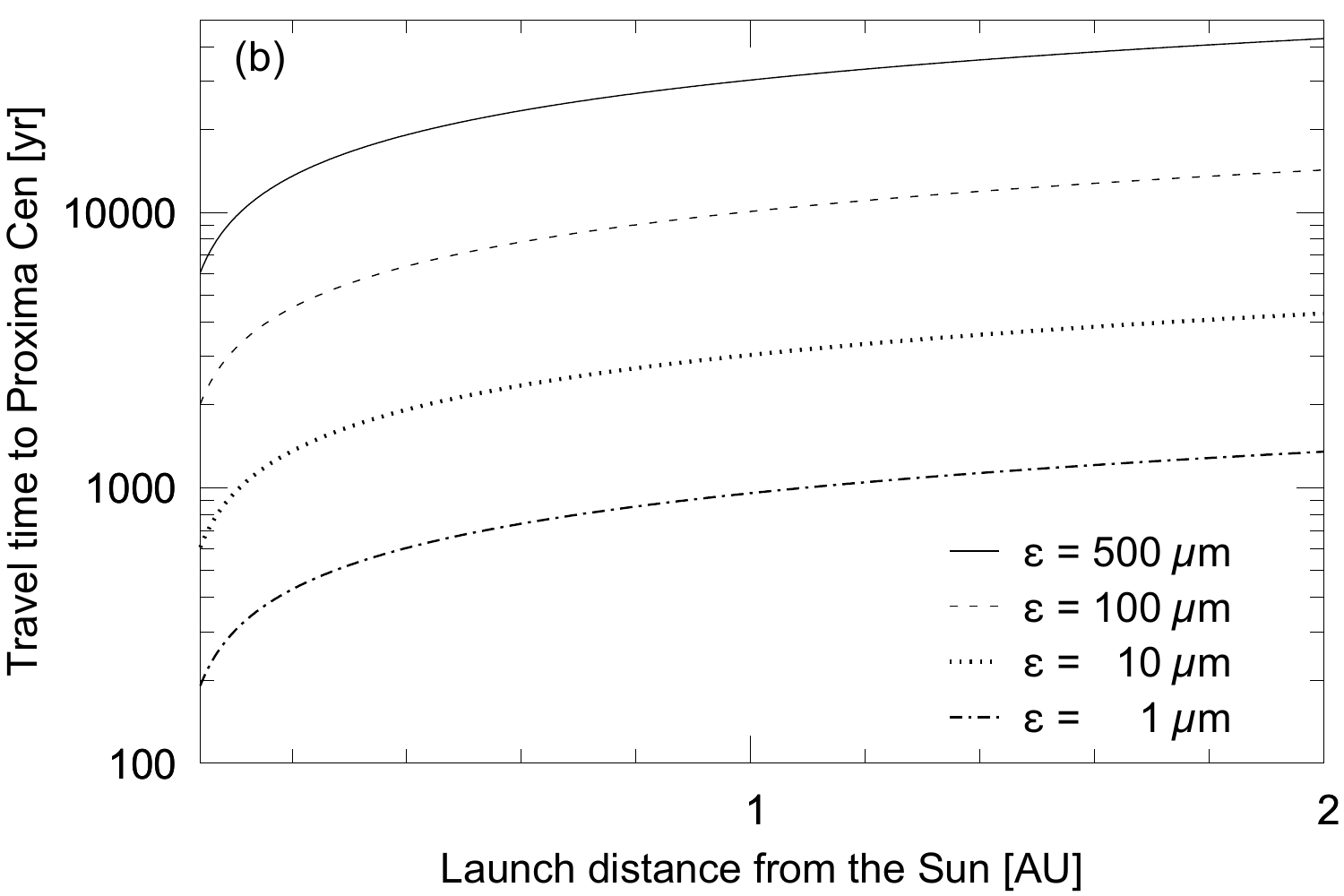}
\caption{Travel characteristics of an aerographite solar radiation-driven hollow sphere. Tracks for four different choices of  shell thickness $\epsilon$ are shown. (a) Terminal speed at infinite distance from the Sun as a function of the launch distance from the Sun. Results were computed with Eq.~\eqref{eq:escape}. (b) Travel time to the nearest star. }
\label{fig:travel}
\end{figure*}

 \begin{figure*}[t]
\centering
\includegraphics[angle= 0, width=1\linewidth]{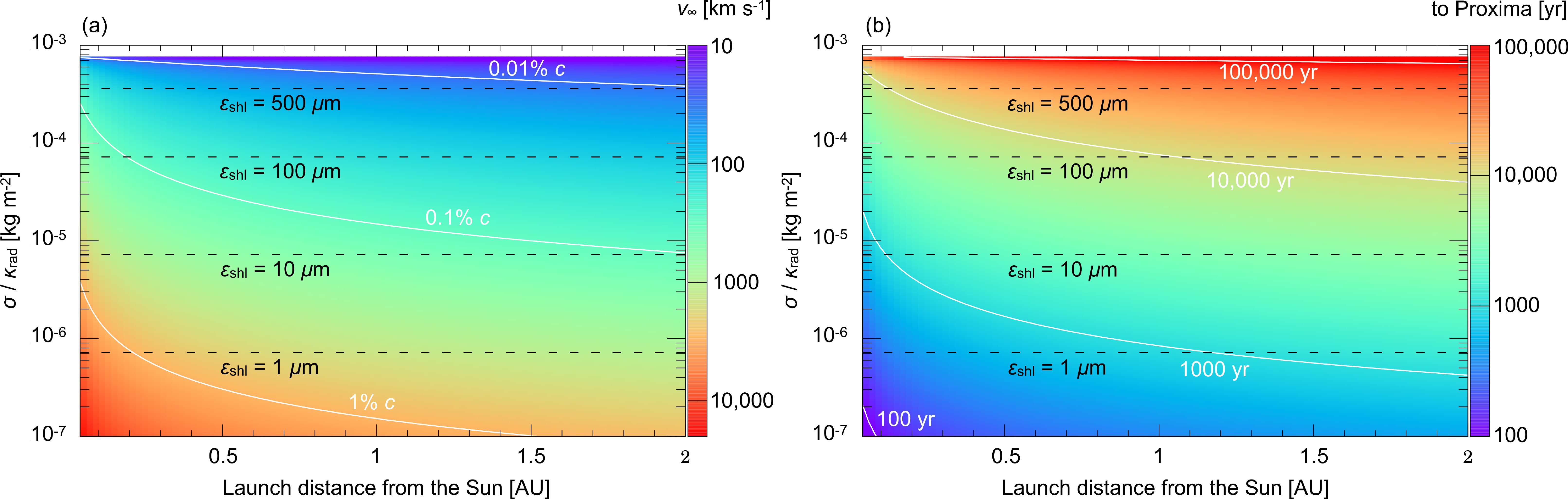}
\caption{Travel characteristics of a solar sail. Results are shown as per Eq.~\eqref{eq:escape} and as a function of the launch distance from the Sun (abscissa) and mass per cross section ratio over the radiative coupling constant (ordinate). Horizontal dashed lines refer to an aerographite hollow sphere with shell thickness values corresponding to those chosen in Fig.~\ref{fig:travel}. (a) Terminal speed at infinite distance from the Sun. White contours show $v_\infty$ in multiples of the speed of light (see white labels). (b) Travel time to the nearest star, Proxima Centauri. White contours refer to constant travel times (labeled in white).}
\label{fig:travel_2}
\end{figure*}

\subsection{Terminal speed}

The terminal velocity ($v_\infty$) of the sail is a key feature to interstellar travel. Although it can be calculated from the numerical simulations outlined in the previous subsection, there is  a more elegant way. We use the conservation of energy to compute the terminal speed ($v_\infty$) of a sail driven by the solar radiation, assuming that the sail has to overcome the solar gravitational potential. The kinetic energy of the sail at a distance $r$ from the Sun is $E_{\rm kin}~=~mv(r)^2/2$.

The resulting force is central (see Eq.~\ref{eq:totalforce});  it only depends on the radial distance to the Sun and it is conservative, which allows us to assume conservation of the total energy $E(r)$. At a finite distance $r$ from the Sun, we have

\begin{align} \label{eq:potential_LEO} \nonumber
V(r) =& \int\displaylimits_{\infty}^{r} {\rm d}r \, F(r) = \int\displaylimits_{\infty}^{r} {\rm d}r \frac{L_\odot}{4 \pi c r^2} S \kappa_{\rm rad} - \frac{GM_\odot m}{r^2} \\
    =& \frac{1}{r} {\Bigg (} \frac{L_\odot}{4 \pi c} S \kappa_{\rm rad} - G M_\odot m {\Bigg )}  + C.
\end{align}

\noindent
The integral over the force requires an integration constant $C$,  chosen such that the energy at infinity of a particle with zero velocity becomes $0$. As a consequence, the total energy at $r=\infty$ for a particle with $v_\infty~\equiv~v(r=\infty)$ collapses to the kinetic term $E_{\infty} = m v_{\infty}^2/2$. With $E(r)=E_{\rm kin}(r)+V(r)$ energy conservation $E(r)=E_\infty$ is equivalent to $E_{\rm kin}(r)+V(r)=E_\infty$, which translates into

\begin{eqnarray}
\frac{1}{2} m v_{\infty}^2 = m \, {\Bigg (} \frac{1}{2} v^2(r) + \frac{1}{r} {\Big (} \frac{L_\odot}{4\pi c} \frac{\kappa_{\rm rad}}{\sigma} - G M_\odot {\Big )} {\Bigg )}
,\end{eqnarray}

\noindent
which is equivalent to

\begin{eqnarray} \label{eq:escape}
v_{\infty} = \sqrt{ v^2(r) + \frac{1}{r} {\Bigg (} \frac{L_\odot}{2\pi c} \frac{\kappa_{\rm rad}}{\sigma} - 2G M_\odot {\Bigg )} } 
.\end{eqnarray}

\noindent
As an aside, for initial zero velocities $v(r)=0$ and neglecting the solar radiation, Eq.~\eqref{eq:escape} can be used to derive the solar system escape velocity from the surface of the Sun ($617.5\,{\rm km\,s}^{-1}$) for $r=R_\odot$ (the solar radius) or from the orbit of the Earth ($42.1\,{\rm km\,s}^{-1}$) for $r=1$\,AU.

In Fig.~\ref{fig:travel}(a), we show $v_\infty(r)$ for an aerographite hollow sphere with $\sigma~=~\sigma_{\rm shl}~=~4\rho\epsilon$ (Eq.~\ref{eq:shell}), $\rho~=~0.18\,{\rm kg\,m}^{-3}$, and shell thicknesses of $0.5\,\epsilon_{\rm shl,aer}'~=~500\,\mu$m (solid line), $0.1\,\epsilon_{\rm shl,aer}'~=~100\,\mu$m (dashed line), $0.01\,\epsilon_{\rm shl,aer}'~=~10\,\mu$m (dotted line), and $0.001\,\epsilon_{\rm shl,aer}'~=~1\,\mu$m (dash-dotted line). For the radiative coupling constant we chose $\kappa_{\rm rad}~=~1$. At $r~=~1$\,AU the values range between about $42\,{\rm km\,s}^{-1}$ for $\epsilon~=~500\,\mu$m and approximately $1331\,{\rm km\,s}^{-1}$ for $\epsilon~=~1\,\mu$m. For a launch distance of 0.04\,AU (closest approach planned for the Parker Solar Probe) the terminal speed increases to about $211\,{\rm km\,s}^{-1}$ ($\epsilon~=~500\,\mu$m) or about $6657\,{\rm km\,s}^{-1}$ ($\epsilon~=~1\,\mu$m), respectively. In Fig.~\ref{fig:travel}(b), we illustrate the resulting travel times to Proxima\,b, which range between 956\,yr ($\epsilon~=~1\,\mu$m) and 30,204\,yr ($\epsilon~=~500\,\mu$m) assuming launch from $r~=~1$\,AU. Launch from as close as 0.04\,AU to the Sun would lead to travel times as short (cosmologically speaking) as 191\,yr ($\epsilon~=~1\,\mu$m) to 6041\,yr ($\epsilon~=~500\,\mu$m), respectively.

Figure~\ref{fig:travel_2} shows a different perspective on these travel characteristics, now as a contour plot over the launch distance from the Sun (abscissa) and $\sigma/\kappa_{\rm rad}$. Panel (a) is a color-coded illustration of $v_\infty(r)$, with a color-to-speed conversion shown in the color bar. The ordinate refers to the material properties of the sail. Four horizontal dashed lines are chosen as examples to connect this visualization to Fig.~\ref{fig:travel}, again featuring a hollow sphere sail design with four different shell thicknesses as discussed above. Panel (b) shows the resulting travel times to Proxima Centauri.

The range of terminal speeds derived for our choices of a plausible sail material, shape, and composition are in agreement with the projection of the historical speed development of man-made vehicles into the near future. Historical speed records can be nicely described by a speed doubling law, or in a relativistic framework by a kinetic energy growth law. For the year 2040 the projected maximum speed achieved by humanity is $0.01\,\%\,c$ \citep{2017MNRAS.470.3664H}, which interestingly is very close to the terminal speed that can be obtained by a $500\,\mu$m thick aerographite sphere after launch at 1\,AU from the Sun, as shown in Fig.~\ref{fig:travel_2}.

\section{Interstellar escape from low-Earth orbit}
\label{sec:Interstellar_LEO}

\subsection{Unbound condition}
\label{sec:unbound}

 \begin{figure*}[t]
\centering
\includegraphics[angle= 0, width=.49\linewidth]{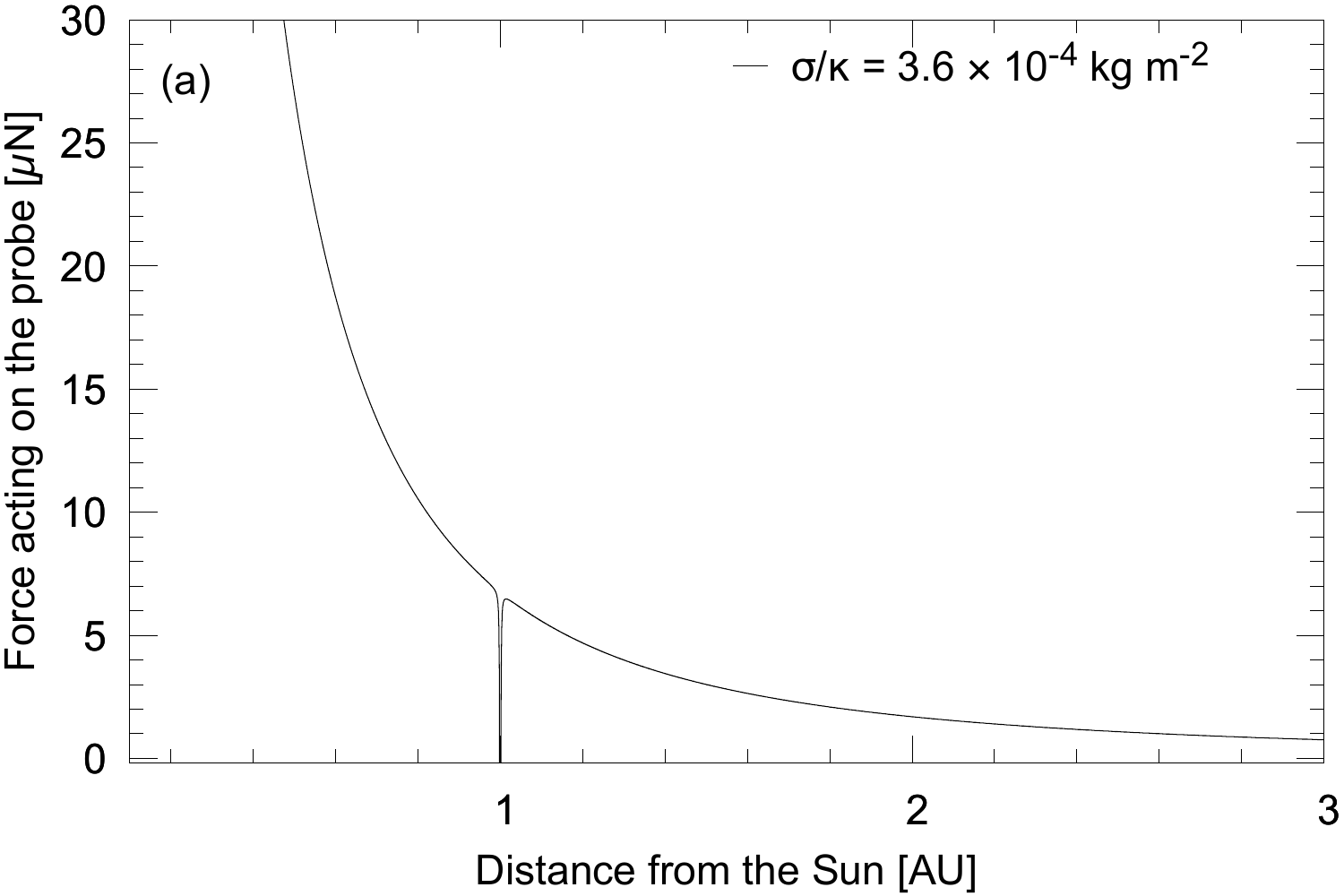}
\includegraphics[angle= 0, width=.49\linewidth]{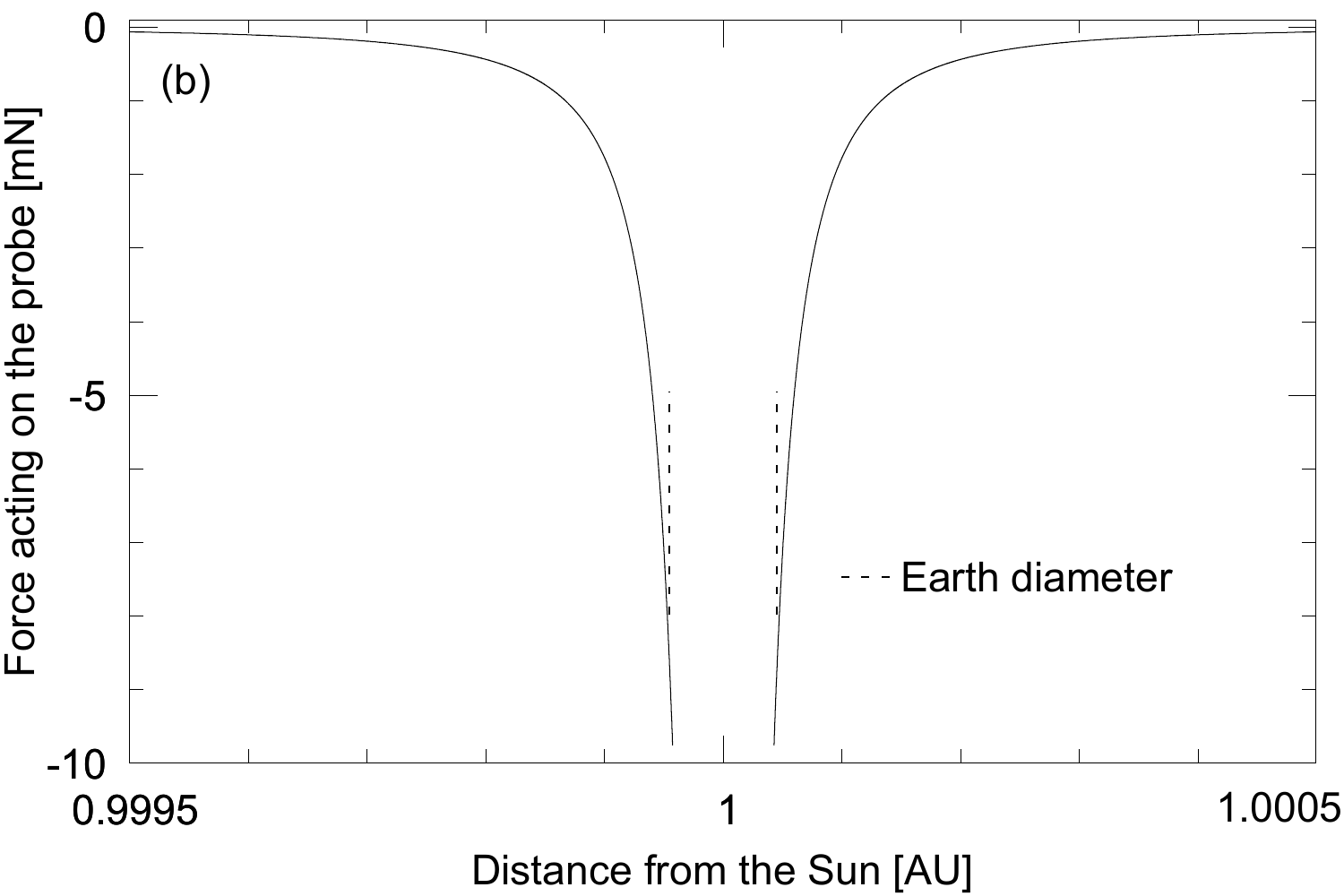}
\caption{Total force resulting from the attractive
gravitational field of the Sun and the Earth, and the repulsive force from the solar radiation. The sail is assumed to have a mass of $m~=~1$\,g, a radiative coupling constant of $\kappa_{\rm rad}=1$, and a mass per cross section area ratio of $\sigma~=~3.6~\times~10^{-4}\,{\rm kg\,m}^{-2}$, about a factor of two below the critical value of $\sigma_\odot~=~7.6946 \times 10^{-4}\,{\rm kg\,m}^{-2}$ to enable interstellar escape from interplanetary space. For a 1\,g aerographite hollow sphere ($\sigma~=~4\rho\epsilon~=~m/S$) with a shell thickness of $\epsilon~=~500\,\mu$m this corresponds to a cross section area of 2.78\,m$^2$ and a radius of 94\,cm. (a) Throughout this 1D cut through the solar system, $F_{\rm tot}~>~0$ because $\sigma/\kappa_{\rm rad}~<~\sigma_\odot$, except in the vicinity of the Earth at 1\,AU. Units along the ordinate are ${\mu}$N. (b) Within a fraction of an AU around the Earth, the Earth's gravitational well is too deep for this sail to escape and $F_{\rm tot}~<~0$. Units along the ordinate are mN.}
\label{fig:forces}
\end{figure*}

So far we have considered the force balance to derive conditions for an escape from {interplanetary} space to interstellar space. Now we investigate the force balance under the additional impression of the Earth's gravitational field. In this one-dimensional problem, $d=|r-1\,{\rm AU}|~>~R_\oplus$ is the distance of the sail from the center of the Earth. The total force acting on the sail then is

\begin{align} \nonumber
F_{\rm tot}(r) = & \ \frac{L_\odot}{4 \pi c r^2} S \kappa_{\rm rad} - \frac{GM_\odot m}{r^2} - \frac{G M_\oplus m}{d^2} \\ \label{eq:Ftot}
                   = & \ m {\Bigg (} \frac{L_\odot}{4 \pi c r^2} \frac{\kappa_{\rm rad}}{\sigma} - G {\Big (} \frac{M_\odot}{r^2} + \frac{M_\oplus}{d^2} {\Big )} {\Bigg )}.
\end{align}

\noindent
For our illustration of Eq.~\eqref{eq:Ftot} in Fig.~\ref{fig:forces}, we make use of the relation $r~=~1\,{\rm AU}+d$ so that the total force becomes a function of the Earth--sail distance, $F_{\rm tot}(d)$. Figure~\ref{fig:forces} assumes a sail with $\kappa_{\rm rad}=1$ and $\sigma~=~3.6~\times~10^{-4}\,{\rm kg\,m}^{-2}$. For a $500\,\mu$m thick  aerographite hollow sphere sail with a mass of $m=1$\,g this implies $S=2.78\,{\rm m}^2$ through Eq.~\eqref{eq:shell}.

We use Eq.~\eqref{eq:Ftot} to derive a critical mass per cross section area for a sail to become unbound from the combined Sun--Earth gravitational potential using the solar photon flux alone. We require that $F_{\rm tot}(d)~>~0$, which is equivalent to

\begin{equation} \label{eq:sigma_LEO}
\frac{\sigma}{\kappa_{\rm rad}} < \frac{L_\odot}{4 \pi c G {\Big (} M_\odot + \frac{\displaystyle r^2}{\displaystyle d^2} M_\oplus {\Big )} } \ .
\end{equation}

\noindent
Substitution of $r~=~1\,{\rm AU}+d$ in our one-dimensional model defines

\begin{equation} \label{eq:sigma_ear}
\sigma_\oplus(d) \equiv \frac{L_\odot}{4 \pi c G {\Big (} M_\odot + \frac{\displaystyle (1\,{\rm AU}+d)^2}{\displaystyle d^2} M_\oplus {\Big )} } \ ,
\end{equation}

\noindent
which shows that Eqs.~\eqref{eq:sigma_LEO} and \eqref{eq:sigma_ear} (using $\kappa_{\rm rad}~=~1$) converge to Eq.~\eqref{eq:conditionSigma} (the critical surface density for solar system escape from interplanetary space) in the limit of $M_\odot~\gg~M_\oplus(1\,{\rm AU}+d)^2/d^2$ or, equivalently, $d~\gg~1\,{\rm AU}/(\sqrt{M_\oplus/M_\odot} - 1)~\sim~41\,R_\oplus$.

 \begin{figure}[t]
\centering
\includegraphics[angle= 0, width=1\linewidth]{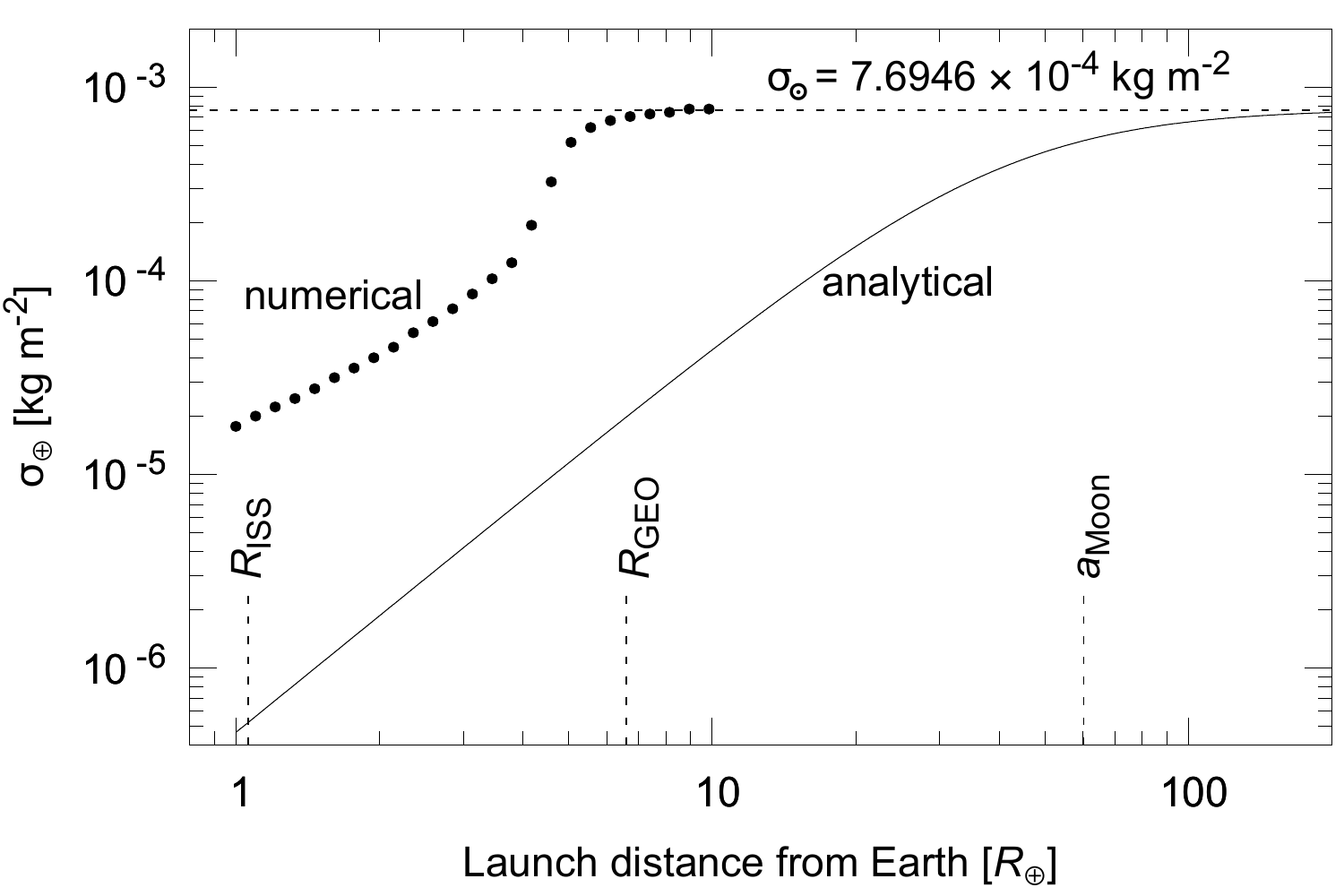}
\caption{Critical mass per cross section of a photon sail to become gravitationally unbound from the solar system after launch in the vicinity of the Earth. The solid line shows values computed with Eq.~\eqref{eq:sigma_ear}. The dots show results from our 2D numerical integrations. Vertical dashed lines indicate the orbit of the International Space Station (400\,km above ground; $1.06\,R_\oplus$ from the Earth's center), the geostationary orbit (altitude of 35,786\,km; $6.6\,R_\oplus$ from the Earth's center), and the orbit of the Moon (378,021\,km above the ground; $60\,R_\oplus$ from the Earth's center).}
\label{fig:sigma_ear}
\end{figure}

 \begin{figure*}[t]
\centering
\includegraphics[angle= 0, width=.495\linewidth]{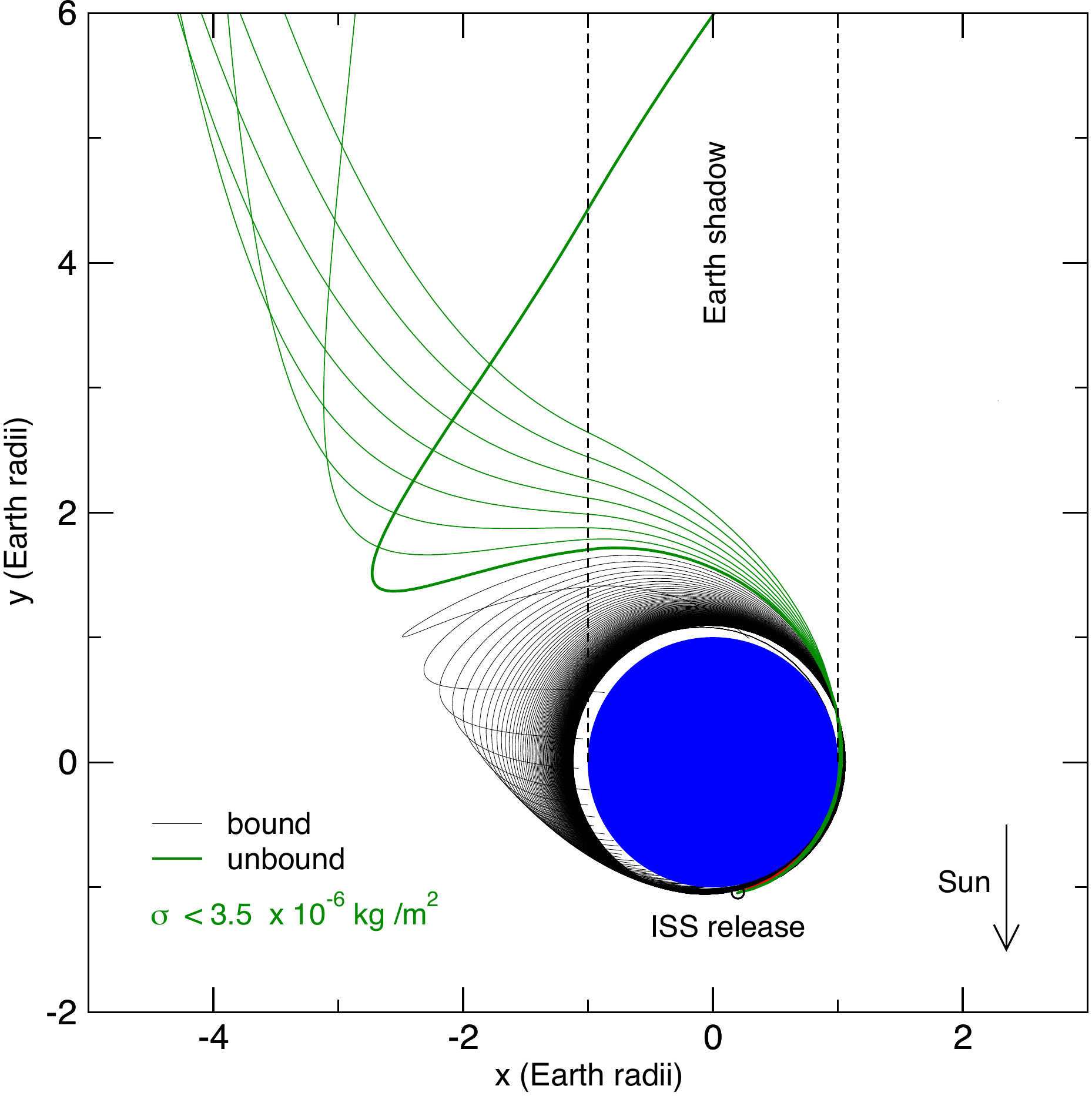}
\hspace{0.2cm}
\includegraphics[angle= 0, width=.482\linewidth]{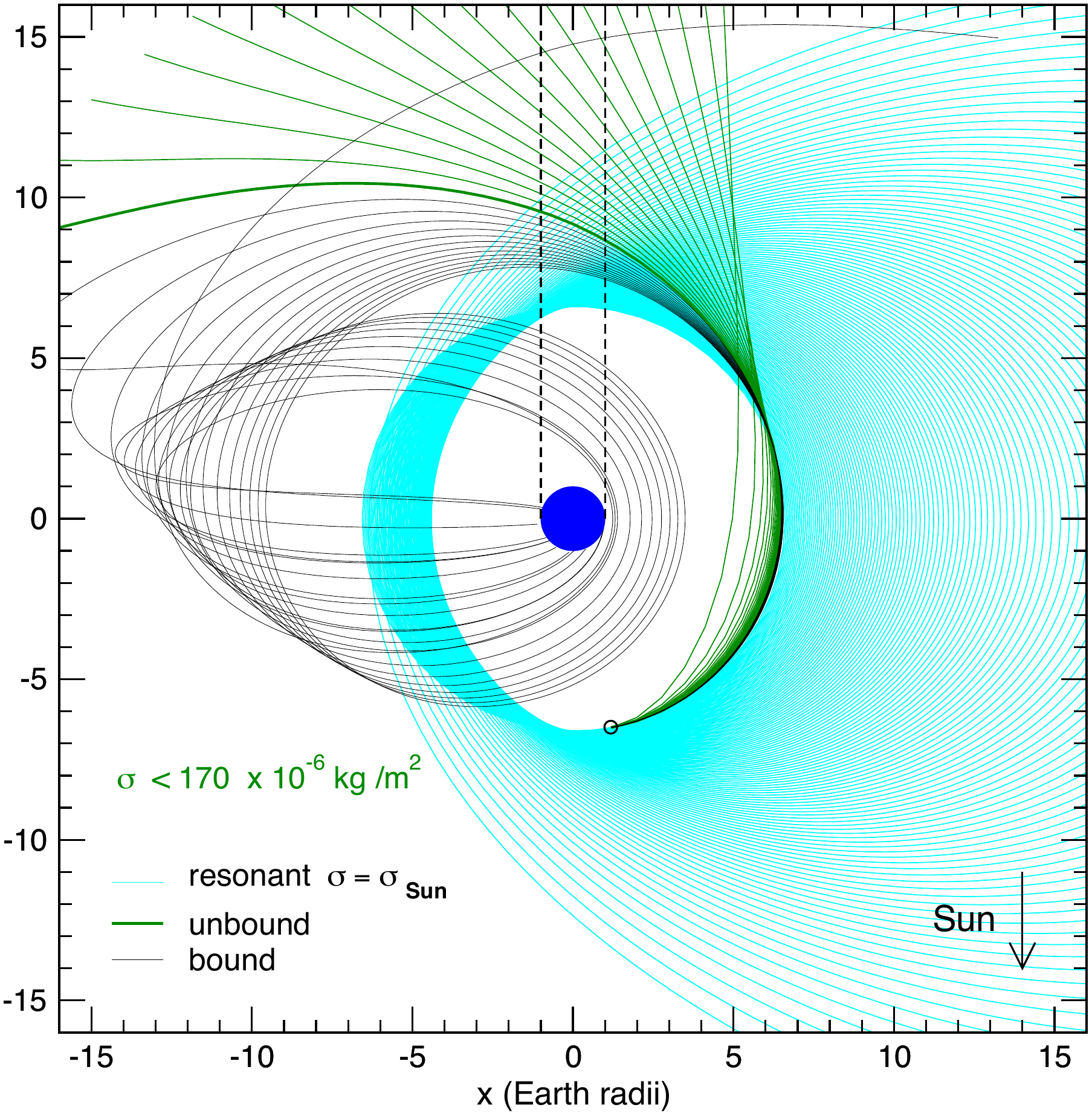}
\caption{Numerical simulations of an aerographite sail under the effect of the solar radiation force and the gravitational field of the Earth and the Sun. All trajectories assume $\kappa_{\rm rad}~=~1$. The Earth is shown as the large filled blue circle. The radiative force is switched off in the Earth's shadow (dashed lines). The direction of the Sun is indicated with an arrow. \textit{Left}: Launch from ISS. Black lines illustrate bound orbits resulting in collision with the Earth. Green lines show successful launches into interstellar space with sufficiently low $\sigma$ values (see legend). Red lines indicate immediate collision with Earth after release from ISS. The step size for $\sigma$ in these numerical simulations is $0.25~\times~10^{-6}\,{\rm kg\,m}^{-2}$. \textit{Right}: Launch from geostationary orbit. Line styles correspond to those in the left panel. The cyan lines (four in this plot) refer to the resonance for $\sigma$ close to $\sigma_\odot$. The step size for $\sigma$ in these numerical simulations is $2.5~\times~10^{-6}\,{\rm kg\,m}^{-2}$.}
\label{fig:trajectories}
\end{figure*}

Equation~\eqref{eq:sigma_ear} reveals that the critical mass per cross section ratio depends on the distance between the sail and the Earth. In other words, the farther away the sail can be launched from Earth, the more massive it can be in relation to its cross section with respect to the solar photon pressure and still achieve escape into interstellar space. This dependency of $\sigma_\oplus~=~\sigma_\oplus(d)$ is fundamentally different from $\sigma_\odot$, which does not depend on $r$ and it is rooted in the fact that the radiation emanates from the Sun.

This behavior is seen in Fig.~\ref{fig:sigma_ear}, where we show $\sigma_\oplus(d)$ (solid line) with $d$ shown in units of Earth radii along the abscissa. The function $\sigma_\oplus(d)$ is only defined for $d~>~R_\oplus$, that is, above the Earth's surface. At that point, we find $\sigma_\oplus(R_\oplus)~=~4.65~\times~10^{-7}\,{\rm kg\,m}^{-2}$. If the Earth had no atmosphere and any other effects could be neglected as well, then a particle with a mass per cross section ratio smaller than $\sigma_\oplus(R_\oplus)$ would be blown into interstellar space by sunlight. At an altitude of 400\,km, roughly corresponding to the orbit of the International Space Station, we find $\sigma_\oplus(R_{\rm ISS})~=~5.26~\times~10^{-7}\,{\rm kg\,m}^{-2}$.

The solid line in Fig.~\ref{fig:sigma_ear} shows how $\sigma_\oplus(d)$ increases by more than three orders of magnitude from $d~=~R_\oplus$ to the orbit of the Moon at about $60\,R_\oplus$, where it reaches $\sigma_\oplus(a_{\rm Moon})~=~5.28~\times~10^{-4}\,{\rm kg\,m}^{-2}$. Ultimately, $\sigma_\oplus(d)$ converges to $\sigma_\odot~=~7.6946~\times~10^{-4}\,{\rm kg\,m}^{-2}$ for large distances from Earth.

\subsection{Numerical integration of the force equation}

The force balance equations in Sect.~\ref{sec:Interstellar_LEO} and the resulting condition on $\sigma_\oplus(d)$ for interstellar escape in Eq.~\eqref{eq:sigma_ear} ignore not only the heliocentric orbital motion, but also the geocentric orbital speed of the sail at the time of launch. If the sail were released at this particular point on its geocentric orbit when it is moving away from the Sun, then the additional speed at this moment of launch would allow for some margin in the ratio $\sigma/\kappa_{\rm rad}$  
and would increase the critical maximum values shown in Fig.~\ref{fig:sigma_ear} possibly by a significant amount.

Our previous calculations in one dimension also ignore the Earth's shadow. If released at the point of maximum tangential deflection from Earth and at the point when moving away from the Sun, we estimate that the sail would have about 5.5\,\% of its ${\sim}90$\,min orbit (about 5\,min) before it  entered the Earth's shadow and no longer be subject to the repulsive solar radiation force. The sail would only be propelled towards outer space again upon egress. At this point, however, its orbital elements would be very different from those at launch.

To address the effects of the geocentric orbital motion at launch as well as the effect of the Earth's shadow in detail, we resort to numerical computations in two spatial dimensions, $\vv{r}~=~(x,y)$. We derive the acceleration of the sail with initial conditions corresponding to a Keplerian orbit from the total force in two dimensions,

\begin{eqnarray}
\frac{d\vv{r}}{dt} &=& \vv{v} \ , \\
\frac{d\vv{v}}{dt} &=& \frac{1}{m} {\Big (} \vv{F}_{{\rm grav},\oplus} + \vv{F}_{{\rm grav},\odot} + \vv{F}_{{\rm rad},\oplus} {\Big )} \ ,
\end{eqnarray}

\noindent
where $\vv{r}$ is the geocentric vector position of the particle, $\vv{v}$ is its geocentric velocity, $\vv{F}_{{\rm grav},\oplus} + \vv{F}_{{\rm grav},\odot}$ is the sum of the gravitational force of the Earth and the Sun, and $\vv{F}_{{\rm rad},\oplus}$ is the radiation pressure force from solar radiation. The purpose of this integration is the qualitative assessment of the typical effect of the dominant forces in the Earth's immediate environment; thus,  second-order effects such as the reflected radiation pressure from Earth and thermal emission, the non-inertial forces due to Earth's orbit around the Sun, and the gravitational effect of the moon and other solar system bodies are not included. The shadow cast by the Earth is included in the simulation by imposing that $\vv{F}_{{\rm rad},\oplus}$ be zero when the Earth is between the particle and the Sun. In addition and  for simplicity, all simulations were performed assuming orbits on the ecliptic plane of the solar system. Perturbations from the Moon's gravity and the Lorentz force exerted by the Earth's magnetic field are ignored, and the motion of the Earth around the Sun is also ignored.

We integrate the equations of motion numerically for one year of simulated time using a constant time step (${\Delta}t$) of 1/1000 of the initial orbital period. The integration is performed using a fourth-order Runge-Kutta integrator \citep{Runge1895,Kutta1901}. Integration is stopped when one of the following conditions is reached: the particle collides with Earth ($r<R_\oplus$), the particle reaches the Earth--Moon distance ($r>57 R_\oplus$), or the particle performs 1000 orbits around the planet.

In Fig.~\ref{fig:trajectories} we show examples for the resulting trajectories. The left panel features a family of trajectories for launch from the ISS in LEO. Black lines refer to trajectories that lead to bound orbits (and ultimate collision with Earth) because the mass per cross section area for these sails is too high. Green trajectories signify interstellar escape, which we find occurs for $\sigma~\leq~3.5~\times~10^{-6}\,{\rm kg\,m}^{-2}$. That said, $\sigma~\leq~1.5~\times~10^{-6}\,{\rm kg\,m}^{-2}$ leads to immediate collision of the sail with the Earth (red lines), and so there is a sensible window of $\sigma$ values suitable for successful escape. We have also sampled the ISS orbit for different launch positions and found that the critical mass per cross section area only weakly depends on the start position. The critical value of  $3.5~\times~10^{-6}\,{\rm kg\,m}^{-2}$ is about a factor of 6.7 higher than our analytical approximation in Sect.~\ref{sec:unbound}.

The right panel of Fig.~\ref{fig:trajectories} illustrates a family of trajectories upon launch from geostationary orbit. Interstellar escape is possible for $\sigma~\leq~1.7~\times~10^{-4}\,{\rm kg\,m}^{-2}$, which is a factor of 8.5 higher than our analytical prediction.

Interestingly, by further increasing $\sigma$ up to values near $\sigma_\odot$, we discover a resonant behavior. A family of four trajectories with $\sigma~\sim~\sigma_\odot$ is shown in cyan lines in the right panel of Fig.~\ref{fig:trajectories}. The sail is forced into a nearly elliptical orbit with apogee beyond $60\,R_\oplus$ irrespective of the launch position along the geostationary orbit. This resonance occurs for all launch altitudes, but the window of $\sigma$ values susceptible to this phenomenon decreases with decreasing launch altitude. For launch altitudes $\lesssim~4\,R_\oplus$ the step size of $\sigma$ used for the trajectories in Fig.~\ref{fig:trajectories} is too small to find these resonances, but we have verified manually that these resonances exist. The maximum  $\sigma$ value for interstellar escape as a function of distance from Earth resulting from our numerical integrations is shown with dots in Fig.~\ref{fig:sigma_ear}.

\section{Follow-up monitoring of the sail}

For the mission to be declared a success and in order to maximize its scientific return, the sail must be tracked as long and far as possible on its journey. Since aerographite is black, it is impossible to track its reflected sunlight.

Instead, a black aerographite sphere could be observed in the infrared (IR). If independent distance  measurements were available, the apparent magnitude could allow studies of interplanetary extinction and offer a new method for the exploration of the interplanetary medium.

Another possibility is through onboard communication equipment. A laser sending the proper time of the sail to Earth would allow distance and speed measurements through the relativistic Doppler effect. Measurements of gravitational perturbations \citep{2017ApJ...834L..20C,2020arXiv200414192W} under consideration of dust and gas drag as well as magnetic forces exerted from the interstellar medium \citep{2020ApJ...895L..35H} could also be used to search for the suspected Planet Nine in the outskirts of the solar system. Its expected orbital semimajor axis is between about 380\,AU and 980\,AU \citep{2016AJ....151...22B,2016ApJ...824L..23B}.

\subsection{Infrared observations from space}

Ground-based IR observations are complicated by the strong absorption of atmospheric water vapor. Instead, spaced-based observations, for example with the James Webb Space Telescope (JWST; operations planned from 2021 to 2031), could allow tracking of the probe. How far out in the solar system could an aerographite sphere be observed? To answer this question we compute its temperature in thermal equilibrium with the absorbed sunlight,

\begin{equation}
T(r) = {\Bigg (} \frac{L_\odot (1-\alpha)}{4 \pi r^2 f \sigma_{\rm SB} } {\Bigg )}^{1/4} \ ,
\end{equation}

\noindent
where $\sigma_{\rm SB}$ is the Stefan-Boltzmann constant, $f~=~2$ is the energy flux redistribution factor for a non-rotating aerographite sphere, and $\alpha~=~0$ is the Bond albedo of aerographite. Next we calculate the radiative intensity (the wavelength-integrated spectral radiance of a black body) of a swarm of $n$ sails,

\begin{equation}\label{eq:I_r}
I(r) = n \ \frac{\sigma_{\rm SB} T(r)^4 S }{ \pi (r-{\rm AU})^2} \ ,
\end{equation}

\noindent
where the nominator (equal to $\pi d^2$) signifies that we consider a one-dimensional radial sail trajectory from the Sun to the Earth and beyond. The bolometric magnitude of the swarm is then given as

\begin{equation}
m_{\rm s}(r) =  -2.5 \log_{10}{\Bigg (}  \frac{I(r)}{F_0} {\Bigg )} \ ,
\end{equation}

\noindent
using $F_0~=~2.518~\times~10^{-8}\,{\rm W\,m}^{-2}$ as reference flux of a zero bolometric magnitude star. Finally, the peak emission wavelength is given through Wien's displacement law: $\lambda_{\rm peak}(r) = b/T(r)$ with $b~=~2.897771955~\times~10^{-3}$\,K\,m.

 \begin{figure}[t]
\centering
\includegraphics[angle= 0, width=1.\linewidth]{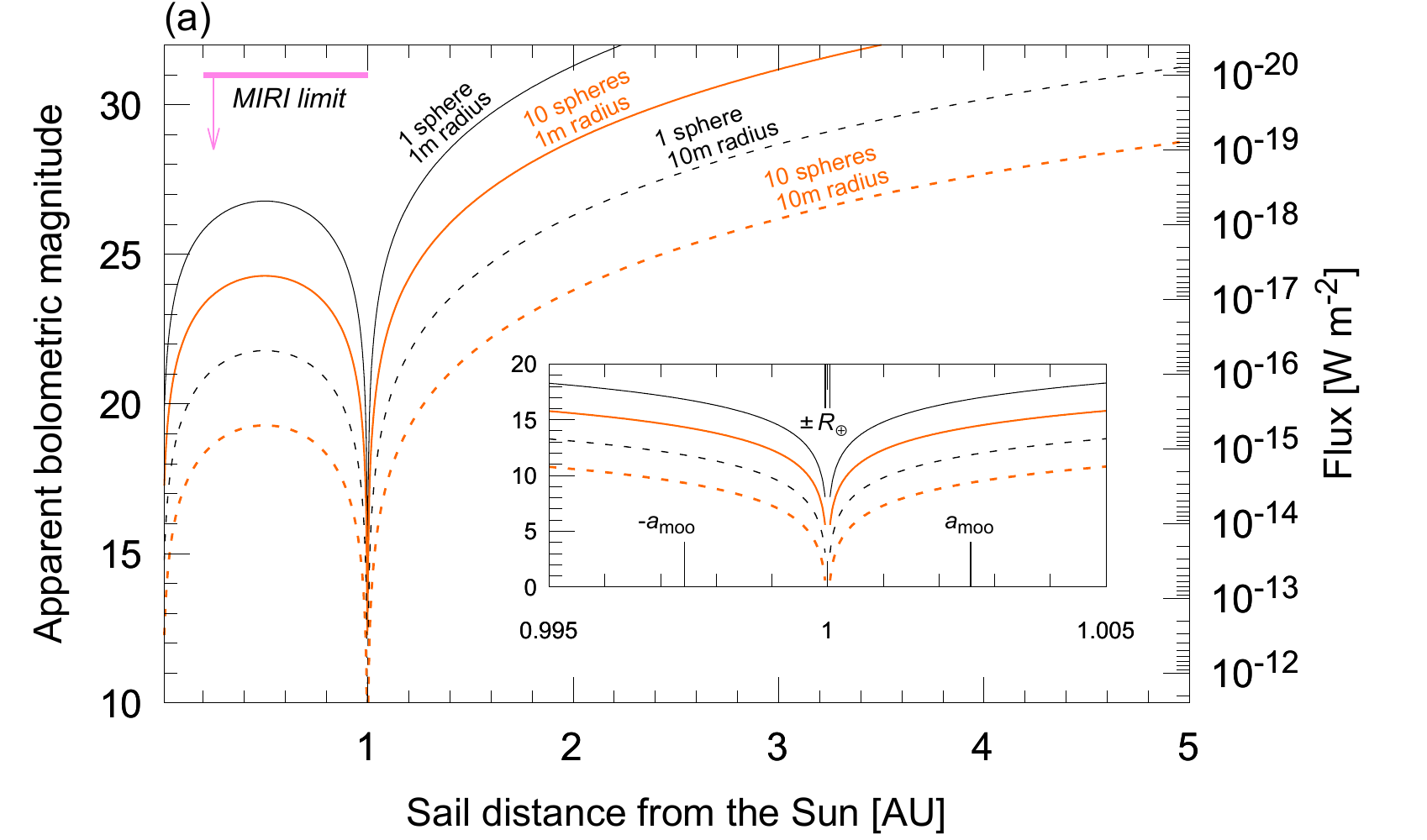}
\includegraphics[angle= 0, width=1.\linewidth]{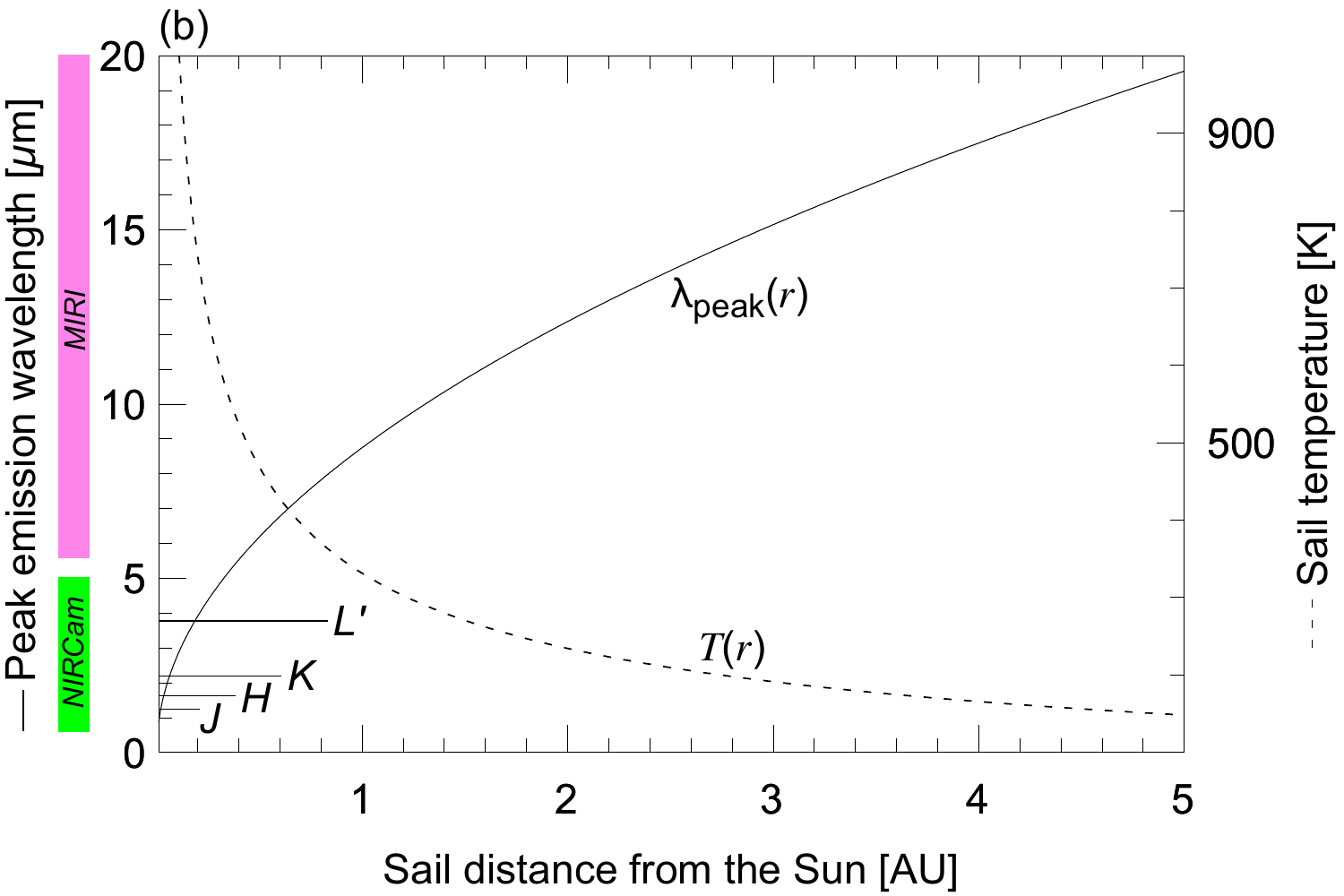}
\caption{(a) Apparent bolometric magnitude (left ordinate) and observed energy flux (right ordinate) of a swarm of spherical solar sails with different radii as a function of distance from the Sun. The observer is supposed to be sitting  1\,AU from the Sun. The $10\,\sigma$ sensitivity limit of a 10\,ks (2.78\,hr) exposure with the MIRI instrument of JWST is indicated at $10^{-20}\,{\rm W\,m}^{-2}$. (b) Peak emission wavelength (left ordinate) and sail temperature (right ordinate) as a function of distance from the Sun. The wavelength coverage of the NIRCam and MIRI instruments of JWST are shown along the left ordinate together with the $JHKL'$ standard filters.}
\label{fig:IR_magnitude}
\end{figure}

Figure~\ref{fig:IR_magnitude}(a) shows $m_{\rm s}(r)$ (left ordinate) and $I(r)$ (right ordinate) for two choices of a sail radius (1\,m and 10\,m) and two choices of a sail swarm size (one and ten), all tracks assuming a hollow sphere (Sect.~\ref{sec:sphere}). Distances $r~<~1$\,AU might be difficult to observe because the sail in this one-dimensional model would be moving towards Earth from the direction of the Sun.

Since Eq.~\eqref{eq:I_r} is proportional to $S~\propto~l^2$, $l$ has a stronger effect on the apparent magnitude than the swarm size, which appears as $n$ in Eq.~\eqref{eq:I_r}. This behavior can be seen in Fig.~\ref{fig:IR_magnitude}(a), where an increase in the sail size from $l~=~1$\,m to $l~=~10$\,m decreases the apparent magnitude at any given distance twice as much as an increase in the swarm size from 1 to 10 objects. That said, an aerographite hollow sphere of 10\,m radius might be challenging to construct let alone to be lifted into space and released from a rocket, satellite, or spaceship. A swarm of many small objects might be more practical to construct, and redundancy would have the benefit of allowing for the loss of some of the objects without total mission failure.

The sensitivity threshold of MIRI of about $10^{-20}\,{\rm W\,m}^{-2}$ \citep{2015PASP..127..686G,2016SPIE.9910E..16P}\footnote{We used the JWST exposure time calculator at \href{https://jwst-docs.stsci.edu}{https://jwst-docs.stsci.edu}.} is shown in the upper left corner of Fig.~\ref{fig:IR_magnitude}(a). Details of a particular observation would depend on the actual trajectory of the sail and in particular on its tangential and radial velocity with respect to JWST. These aspects determine the longest plausible exposure time before the sail would start to smear over several pixels.

As the sail recedes from the Sun, its effective temperature drops and so the wavelength of peak emission increases. This effect is shown in Fig.~\ref{fig:IR_magnitude}(b) with $\lambda_{\rm peak}(r)$ plotted on the left ordinate and $T(r)$ shown on the right ordinate. Also shown are the wavelength coverage of the NIRCam and MIRI instruments of JWST that can be used for standard imaging observations and the $JHKL'$ filters as implemented at the 2.2\,m telescope of ESO at La Silla \citep{1990MNRAS.247..624A}.

JWST observations of an aerographite sail of 1\,m radius would be possible out to about 2\,AU from the Sun, almost midway between the orbits of Mars (1.52\,AU) and of the most massive asteroid Ceres (2.77\,AU). At that distance it would have a temperature of 234\,K and a peak emission wavelength of $12.4\,\mu$m, central to the wavelength coverage of MIRI ($5.6\,\mu{\rm m}~\leq~\lambda~\leq~25.5\,\mu{\rm m}$). Alternatively, a sail 10\,m in radius would be observable out to 3\,AU from the Sun, where it would have cooled to 191\,K with a peak emission at $15.1\mu$m. At any given distance, a swarm of $n$ sails in close formation would decrease the apparent magnitude by $-2.5\,\log_{10}(n)$.

 \begin{figure*}[t]
\centering
\includegraphics[angle= 0, width=1\linewidth]{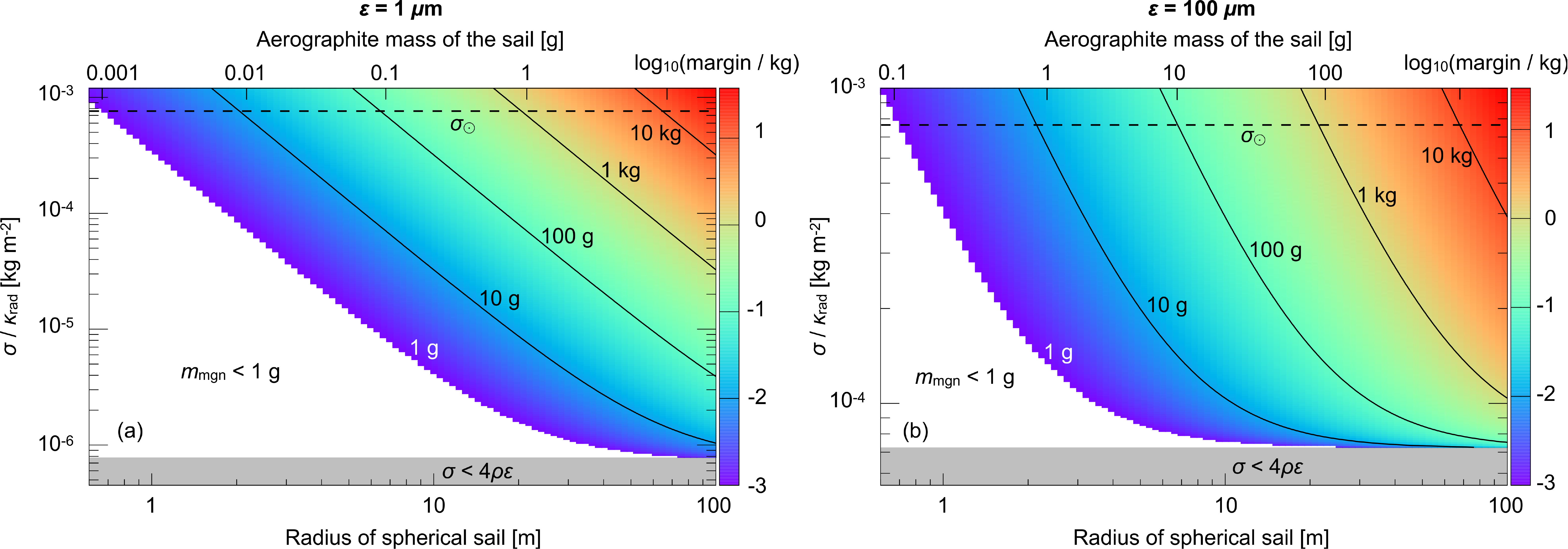}
\caption{Mass margins for a hollow sphere with a shell made of aerographite ($\rho=0.18\,{\rm kg\,m}^{-3}$). Colors indicate $\log_{10}(m_{\rm mgn})$ in  grams (see color scale) whenever the margin is $\geq~1$\,g. The critical value for interstellar escape from interplanetary space in the solar system ($\sigma_\odot~=~7.6946 \times 10^{-4}\,{\rm kg\,m}^{-2}$ as per Eq.~\ref{eq:sigma_I}) is indicated with a dashed line. The gray area indicates the region in which there is no margin for additional mass (see Eq.~\ref{eq:m_mgn}). The abscissa at the bottom shows the sail radius and the abscissa at the top shows the corresponding aerographite mass of the sail without payload. (a) The shell is assumed to have a thickness of $\epsilon~=~1\,\mu$m. (b) The shell is assumed to have a thickness of $\epsilon~=~100\,\mu$m.}
\label{fig:margin}
\end{figure*}

\subsection{Mass margins for onboard equipment}

We now consider the option that the sail carries its own onboard electronic devices such as a laser to communicate with Earth \citep{2016JBIS...69...40L,2020arXiv200508940P}. More fundamental than the question of the content and the design of the signal is the question about the mass margins ($m_{\rm mgn}$) set by the force equation.

Miniaturization of electronic components has made great progress in the last few decades, but we focus on mass margins above 1\,g because we do not expect sub-gram margins to be relevant for the foreseeable future. Commercial lithium-ion batteries weighing a few grams and with power densities $>~1\,{\rm kW\,kg}^{-1}$ \citep{Duduta2018} as well as ultra high-energy density supercapacitors with power densities of ${\sim}32\,{\rm kW\,kg}^{-1}$ \citep{Rani2019} are already available, allowing energy emission of a gram-sized power source of $32$\,W in theory.

First, we express the total mass of a hollow sphere as $m~=~m_{\rm shl}+m_{\rm mgn}$, where $m_{\rm shl}$ is the pure mass of the shell. Then we substitute the expression for $m$ in Eq.~\eqref{eq:shell} and find

\begin{equation}\label{eq:m_mgn}
m_{\rm mgn} = S(\sigma - 4\rho \epsilon) \ ,
\end{equation}

\noindent
where $\sigma~>~4 \rho \epsilon$ is required.

In Fig.~\ref{fig:margin} we plot the resulting mass margin for an aerographite hollow sphere with two different choices of a shell thickness: (a) $\epsilon~=~1\,\mu$m and (b) $\epsilon~=~100\,\mu$m. The dashed line at $\sigma_\odot~=~7.6946 \times 10^{-4}\,{\rm kg\,m}^{-2}$ sets the limit for interstellar escape (see Eq.~\ref{eq:sigma_I}). In panel (a) for $\epsilon~=~1\,\mu$m we find that a 1\,g mass margin would permit interstellar escape (with terminal speed close to zero) of a sphere with a radius of at least 0.65\,m. The weight of an aerographite sphere of this radius is 1\,mg, which means that the mass of the payload is a factor of 1000 greater than the mass of the spacecraft, $m_{\rm mgn}/m~\sim~1000$. For comparison, interstellar missions on chemical rockets such as Voyager~1 on a Titan\,IIIE rocket and New Horizons on an Atlas\,V rocket typically achieve $m_{\rm mgn}/m~\sim~1/1000$.

A sphere with a $2$\,m radius could carry 10\,g in addition to its mere aerographite structure of 10\,mg weight and still achieve interstellar escape with zero terminal speed. As an alternative, a 1\,m (5\,m) radius hollow sphere would weigh 2.3\,mg (57\,mg) and have a margin of 2.4\,g (60\,g) to go interstellar. If it were equipped with a 1\,g (10\,g) load it would have $\sigma~=~3.2~\times~10^{-4}\,{\rm kg\,m}^{-2}$ ($\sigma~=~1.3~\times~10^{-4}\,{\rm kg\,m}^{-2}$). Our numerical simulations show that this results in a terminal speed of about 0.017\,\%\,$c~=~51\,{\rm km\,s}^{-1}$ (0.031\,\%\,$c~=~93\,{\rm km\,s}^{-1}$), or about 3 times (5.5 times) the terminal speed of Voyager~1 if launched in interplanetary space at 1\,AU from the Sun (see Fig.~\ref{fig:travel_2}). At that speed, it would reach the orbit of Pluto after 3.9 (2.1\,yr) of interplanetary travel.

In Fig.~\ref{fig:margin}(b) for a sphere with $\epsilon~=~100\,\mu$m we find again that mass margins above 1\,g for interstellar escape are possible for sphere radii $>~0.65$\,m. Different from panel (a), the pure aerographite mass would contribute about 0.1\,g for this sail size. For any sail radius the aerographite mass is higher by a factor of 100 compared to panel (a), which is determined by the proportionality of $\sigma_{\rm shl}~\propto~\epsilon$ in Eq.~\eqref{eq:shell}. But since the mass of the sail is completely dominated by the payload (and not  the aerographite structure), the travel characteristics for the $\epsilon~=~100\,\mu$m sail are similar to the $\epsilon~=~1\,\mu$m sail if $\sigma/\kappa_{\rm rad}~\sim~\sigma_\odot$. A sail with a 1\,m (5\,m) radius has a mass of 0.23\,g (5.7\,g) and a mass margin of 2.2\,g (55\,g) to go interstellar. A payload of 1\,g (10\,g) would result in $\sigma~=~3.9~\times~10^{-4}\,{\rm kg\,m}^{-2}$ ($\sigma~=~2.0~\times~10^{-4}\,{\rm kg\,m}^{-2}$), $v_\infty~=~41\,{\rm km\,s}^{-1}$ ($v_\infty~=~70\,{\rm km\,s}^{-1}$), and  travel time to the orbit of Pluto within 4.7 yr (2.8\,yr).

\section{Discussion}

\subsection{Flight vector and course correction}

The benefit of a spherical design is in its extremely low mass per cross section area in the limit of a thin shell (Sect.~\ref{sec:sphere}). Moreover, a perfect sphere would have perfect photodynamical stability when riding on light \citep{2017ApJ...837L..20M}. That said, even a small localized imperfection of its reflective--absorptive properties could trigger uncontrolled spin. This might not affect its trajectory or observability from Earth, but it could frustrate communication with Earth for which precise aiming of an onboard communicator (e.g., a laser) would be necessary.

Course corrections of light sails can be made by adjusting its pitch and yaw angles. In our case of a maximally  simple mission concept, onboard thrusters are not available for controlled maneuvering. Instead, an intended slight deformation of the sail into a conical shape would enable passive stabilization \citep{kirpichnikov1995planar,van2001solar}. An axisymmetric, conical solar sail has a stable attitude; that is, it moves on a straight path when illuminated by a distant point source. For this stabilizing shape to be effective, the concave side would need to point towards the Sun. The decrease in surface area to mass compared to a flat sail depends on the cone angle. Stable orientation can be achieved for moderately conical shapes and only reduce the cross section surface area per given mass by a few percent compared to a hollow sphere \citep{2014AdSpR..54...72H}. This holds for a wide range of sail parameters (reflectivity, center of mass), imperfections, and perturbations. Even a hollow hemisphere could result in stabilization, which would actually decrease $\sigma$ by a factor of two compared to the hollow sphere concept studied in this paper. Attitude equilibrium can be obtained by proper arrangements of the center of mass of the sail with respect to the center of the light pressure \citep{2014AdSpR..54...72H}.

\subsection{Non-point source effects}

Our model assumes that the Sun acts as a point source of light. This is a viable assumption for launch at 1\,AU from the Sun and beyond. A mission that  first approached the Sun to increase the terminal speed for solar system escape, however, would encounter significant non-point source effects \citep{1990AcAau..22..155M,1990CeMDA..49..249M} upon which Eq.~\eqref{eq:F_rad} would need to be modified to

\begin{equation}\label{eq:F}
F(r) = \frac{\displaystyle L_\star A}{\displaystyle 3{\pi}cR_\star^2} {\Bigg [} 1- {\Big [}1- {\Big (} \frac{\displaystyle R_\star}{\displaystyle r} {\Big )}^2{\Big ]}^{3/2} {\Bigg ]} \ \ .
\end{equation}

\noindent
This would affect the terminal speed and travel time to Proxima\,Cen shown in Fig.~\ref{fig:travel} as well as the total force acting on the sail illustrated in Fig.~\ref{fig:forces}(a).

\subsection{Other physical effects on the sail trajectory}

Quickly after submission to the solar radiation, an aerographite photon sail could become electrically charged by the solar UV radiation or possibly by the solar wind. If launched from LEO, this charge could lead to a deflection of the sail due to the Lorentz forces induced by the Earth's magnetic field. Moreover, an ultra light sail, as envisioned in this study, could be substantially affected by the air resistance in the Earth's upper atmosphere, a phenomenon known as atmospheric drag.

At the Earth's position around the Sun, the solar wind has a number density on the order of $n=10$ particles per cubic meter \citep{Kepko2003} with a median velocity near $v_{\rm sw}=500\,{\rm km\,s}^{-1}$. The particle flux through a sail is $f~=~Sn|v-v_{\rm sw}|$. Assuming that the relative speed between the sail and the solar wind is comparable to $v_{\rm sw}$ the particle flux is $f~=~5~{\times}~10^6\,{\rm m}^{-2}{\rm s}^{-1}$ and the number of absorbed particles is $N~{\sim}~1.6~\times~10^{14}$ after one year. Their cumulative kinetic energy of $E_{\rm kin,sw}=N m_{\rm pro}v_{\rm sw}^2/2~{\sim}~3~{\times}~10^{-2}$\,J ($m_{\rm pro}$ being the mass of the proton) is negligible compared to the kinetic energy of a 1\,g sail traveling at $100\,{\rm km\,s}^{-1}$, for which $E_{\rm kin}=m v^2/2~{\sim}~5~{\times}~10^6$\,J. The kinetic energy absorbed by the sail is smaller than $E_{\rm kin,sw}$ because the absolute value of the relative speed between the sail and the solar wind is smaller than $v_{\rm sw}$ as long as $v~<~1000\,{\rm km\,s}^{-1}$. Even for the highest sail velocities of $10,000\,{\rm km\,s}^{-1}$ considered in this study the absorption of solar wind particles does not contribute to the velocity budget of the sail.

The calculations in Sect.~\ref{sec:design} show that the critical length scale for interstellar escape is comparable to the wavelength of the solar radiation at peak emission near $\lambda~=~500$\,nm. Mie scattering is thus important to properly calculate the reflective and absorptive properties, which we have encapsulated in the radiation coupling constant $\kappa_{\rm rad}$. Proper treatment will affect our results on the terminal speed and travel time by a factor of a few.

\subsection{Cost estimate}

\subsubsection{Manufacturing costs}

Aerographite stands out as a potential material for an interstellar solar sail in many ways. Beyond its ultra lightweight properties, it can also be fabricated in a large variety of macroscopic shapes (on centimeter scale)  \citep{Mecklenburg2012,Garlof2017}. It is completely optically opaque (thus $\kappa_{\rm rad}~=~1$) and superhydrophobic. It recovers completely after compression by 95\,\%, exhibits outstanding mechanical robustness, specific stiffness, and tensile strength as well as high temperature stability and chemical resistance \citep{Mecklenburg2012}. All these things combined make it an ideal material to maintain structural integrity in the presence of strong vibrations expected during rocket launch into space and to survive the high accelerations expected for the solar sail upon release to the solar wind.

The demonstration of 3D printing of  centimeter-sized structures made of graphene aerogel \citep{Zhang2016}, another carbon-based ultra lightweight material that shares many properties with aerographite, suggests that production of meter-sized structures of similar ultra lightweight material is plausible within the next decade or so. The typical price tag of modern commercial aerogel insulation tiles is on the order of 100\,USD\,m$^{-2}$. For comparison, monolayer graphene is commercially available at a cost of approximately 100\,EUR\,cm$^{-2}$. Weight being a main factor for the price tag of commercial products, the extremely low density of aerographite comes with a key advantage for the scalability of our proposed concept to a swarm of sails. Costs for raw materials are negligible on a per-sail basis, and without sensors the costs for development are substantially reduced compared to more complex mission concepts. It therefore seems reasonable to project that meter-sized aerographite hollow spheres with a thickness of $\mu$m could be produced in large numbers for $1000$\,USD or less per piece. This price tag is comparable to the per-sail cost estimate of 100\,USD for the Breakthrough Starshot mission \citep{Loeb2019}.

These estimates come on top of any development costs towards the final product, which may be on the order of one million USD.

\subsubsection{Launch costs}


Rocket launches have become a business that could be leveraged for future co-ride options. The first launch of a SpaceX Falcon Heavy rocket in early 2018 sent a dummy payload of 1300\,kg into an elliptical orbit beyond the orbit of Mars \citep{2018A&G....59b2.11.,Rein2018}. The additional cost (and risk) of transporting a meter-sized solar sail would have been essentially zero. The publicity effect of a live broadcast of an interstellar mission launch using onboard cameras would potentially be massive.

SpaceX plans to take people to Mars \citep{2017NewSc.236....7.,2018cosp...42E3688W}. Many uncrewed missions will precede such a journey, offering ample low-cost possibilities for a lightweight co-payload in the form of a solar sail. The SpaceX Starship spacecraft and Super Heavy rocket, collectively referred to as Starship, will be the world’s most powerful launch vehicle ever developed, with the ability to carry in excess of 100 metric tons to Earth orbit.\footnote{\href{https://www.spacex.com/vehicles/starship/}{www.spacex.com/vehicles/starship}} The additional weight of an aerographite sail is negligible in this context, but the scientific value and public outreach would likely be immense.

Alternatively, the most expensive option would be to purchase a rocket launch at the common market price. A single launch of a Falcon~9 rocket is available\footnote{May 2020 prices: \href{https://www.spacex.com/about/capabilities}{www.spacex.com/about/capabilities}} for $62$\,million\,USD. The upper stage vehicle is able to inject $4020\,$kg into a Mars orbit, more than enough to send a fleet of millions of gram-sized aerographite sails into interplanetary space, at least from a weight perspective.

\section{Conclusions}

We identify aerographite as a practical low-cost, and low-weight material for a meter-sized solar sail to be pushed to interstellar speed by the solar photon pressure. We show that a hollow aerographite sphere with a shell thickness of up to 1\,mm results in sufficiently low mass per cross section for Sun-driven escape to interstellar space from interplanetary space. If such a sail could be lifted out of the Earth's gravitational well prior to submission to the solar radiation pressure (e.g. as a piggyback mission to an interplanetary mission), no onboard or ground-based propellant would be necessary for the sail to go interstellar. A sail with a 1\,m radius could be tracked through its IR remission of absorbed sunlight with JWST out to a distance of about 2\,AU from Earth, that is,  between the orbits of Mars and the most massive asteroid, Ceres.

Alternatively, a thinner aerographite shell could carry ultra lightweight scientific equipment. Our analytical approximations for launch from interplanetary space show that a hollow graphene sphere with a shell thickness of $0.1$\,mm can carry a payload of several grams to the orbit of Pluto within a few years. In a benchmark scenario, a 1\,m radius hollow aerographite sphere with a shell thickness of $1\,\mu$m ($100\,\mu$m) would weigh 2.3\,mg (230\,mg) and have a margin of 2.4\,g (2.2\,g) to go interstellar. Upon release to the solar radiation in interplanetary space at 1\,AU from the Sun, a payload mass of 1\,g would yield a terminal speed of $51\,{\rm km\,s}^{-1}$ ($41\,{\rm km\,s}^{-1}$), which is 3 times (2.4 times) the terminal speed of Voyager~1. The travel time to the orbit of Pluto would be 3.9\,yr (4.7\,yr). An increase in the sail radius to 5\,m would allow payload masses of 10\,g to reach the orbit of Pluto in almost half the time. Although the weight of the payload is extremely small in these benchmark scenarios, it is a thousand times more than the weight of its transport system. For comparison, the transport systems of the Voyager~1 and New Horizons interstellar missions (Titan\,IIIE and Atlas\,V, respectively) were a thousand times heavier than their payloads.

Our concept is scalable in size and numbers. A swarm of aerographite spheres could be constructed by connecting them with carbon nanofibers, the additional weight of which would be small compared to the mass of the aerographite shell. If mutual shading or entanglement can be avoided, then  their combined thrust would allow larger payloads for a fixed travel duration or, vice versa, faster travel of a given payload. A large swarm of sails could be accessible to deep space tracking with ALMA. At a distance of 1000\,AU the sail temperature would decrease to 10\,K and enter ALMA's sensitivity limit near 900\,GHz. Alternatively, a sail with a 1\,m radius could, in principle, be tracked by a reflection of an active radio ping to distances corresponding to several times the Moon's orbital distance around the Earth.

Our numerical simulations of test particles with cross section ratios corresponding to meter-sized, sub-millimeter thick aerographite shells suggest that interstellar speeds can also be achieved from within the Earth's gravitational potential. Launch from the ISS is possible in principle, but would be complicated by the non-negligible effects of the Earth's magnetic field and atmospheric drag. Instead, launch from geostationary orbit, for example as a piggyback mission to a geostationary satellite mission, would be much more practical. We find a secular resonance that allows escape from the Earth's gravitational field from near geostationary orbit (or beyond) with mass per cross section ratios more than one order of magnitude higher than predicted by the analytical solution.

This mission concept looks like an extremely challenging option,  though principally feasible,   from a manufacturing perspective. Commercial ultra light (1\,g) and ultra high-energy density (${\sim}32\,{\rm kW\,kg}^{-1}$) lithium-ion batteries or supercapacitors could, in theory, be implemented in the sail and permit short-burst energy emission of ${\sim}32$\,W from an onboard miniature laser using a gram-sized power source. A swarm of such minimally equipped aerographite spherical sails could be used to study the Planet Nine hypothesis by measuring relativistic Doppler effects of the laser signal emanating from a sail, the frequency of which would be affected by gravitational perturbations of a massive object.


We estimate a total price tag of less than 1000\,USD per sail, development costs of 1 million USD for a prototype sail, and maximum launch costs of 62 million USD. If implemented as a piggyback mission, launch costs could be reduced dramatically. The total cost would be less than 100\,million USD including overhead, possibly near 10\,million USD in a piggyback mission scenario. The key distinction from the Breakthrough Starshot concept with its total cost estimate far beyond 1\,billion USD is the use of freely available sunlight as a propulsion instead of an expensive ground-based laser array.

Three key challenges for such a mission remain to be addressed: (1)  manufacturing  a meter-sized sub-millimeter thick sphere of aerographite; (2)  producing and installing  gram-scale onboard equipment to enable communication with Earth; and (3) transporting  the ultra thin aerographite sphere into interplanetary space while preserving its structural integrity.

\begin{acknowledgements}
This work made use of NASA's ADS Bibliographic Services. RH is supported by the German space agency (Deutsches Zentrum f\"ur Luft- und Raumfahrt) under PLATO Data Center grant 50OO1501. GAE is supported by a MINECO/Spain fellowship grant (RYC-2017-22489) and his contribution to this research was enabled thanks to a Gauss Professorship granted by the Akademie der Wissenschaften zu G{\"o}ttingen.
\end{acknowledgements}

\bibliographystyle{aa} 
\bibliography{ms}

\begin{appendix}

\section{Other star systems}
\label{sec:app_other}

Other star systems have their individual critical mass per cross section ratio ($\sigma_*$), which depends on the stellar luminosity ($L_*$) and mass ($M_*$). We can generalize Eq.~\eqref{eq:sigma_I} as

\begin{equation} \label{eq:sigma_stars}
\sigma_* = \frac{L_*}{4\pi c G M_*} \ .
\end{equation}

\begin{figure}[t]
\centering
\includegraphics[angle= 0, width=1\linewidth]{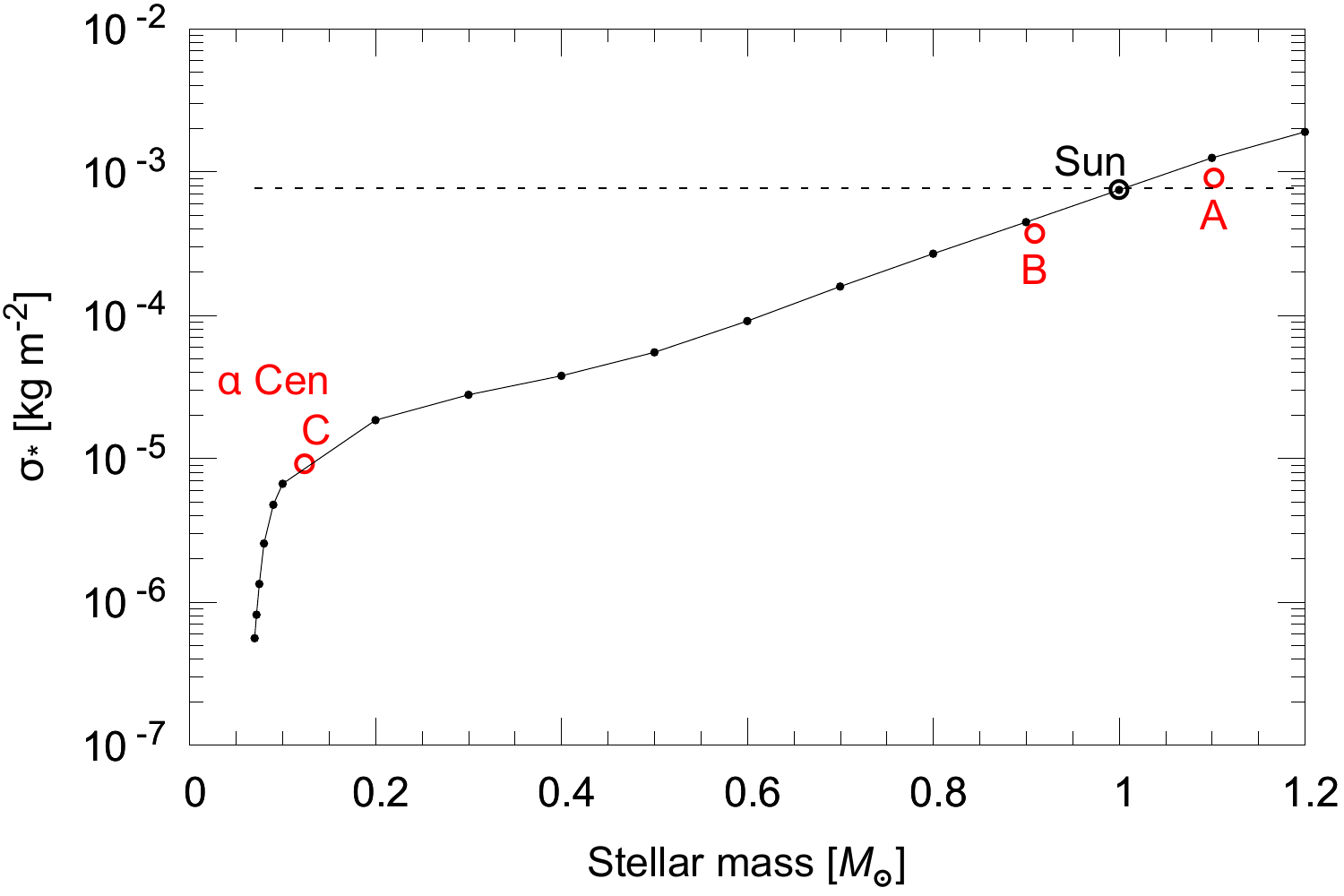}
\caption{Critical mass per cross section ($\sigma_*$) for a photon sail to escape from a main-sequence star into interstellar space. The value of $\sigma_\odot$ from Eq.~\eqref{eq:sigma_I} is  in black. The values of $\alpha$\,Cen\,A, B, and C ($=$\,Proxima\,Cen) are  in red. The dots that are connected by a line use the stellar evolution models of \citet{2015A&A...577A..42B}.}
\label{fig:sigma_stars}
\end{figure}

In Fig.~\ref{fig:sigma_stars} we show $\sigma_*$ for main-sequence stars with masses ranging between $0.07\,M_\odot$ and $1.2\,M_\odot$. The solar value ($\sigma_\odot$) is computed according to Eq.~\eqref{eq:sigma_I} (black label). We also show the values for the three stars of the $\alpha$\,Centauri system for reference (red labels), using Eq.~\eqref{eq:sigma_stars} with mass and luminosity estimates for $\alpha$\,Cen\,A/B from \citet{2002A&A...392L...9T} as well as luminosity \citep{1990A&A...235..335D} and mass estimates \citep{2017A&A...598L...7K} of Proxima\,Cen. The black lined dots cover the entire range of main-sequence stars based on stellar evolution tracks of \citet{2015A&A...577A..42B}, assuming solar metallicity and an age of 4.5\,Gyr.

As a general result, we find that $\sigma_*$ increases with stellar mass. This finding can also be explained without the use of stellar evolution models by invoking the well-known mass--luminosity relation of main-sequence stars. It has long been known that $(L_*/L_\odot)$ can be expressed by a proportionality to $(M_*/M_\odot)^\alpha$, where typically $\alpha$ is much larger than 1 for main-sequence stars \citep{1938ApJ....88..472K}. The results shown in Fig.~\ref{fig:sigma_stars} become relevant in the context of the deceleration of interstellar sails at distant star systems \citep{2017AJ....154..115H} or even return missions \citep{2017ApJ...835L..32H}.

\end{appendix}

\end{document}